 \documentclass[final,5p,times,twocolumn,authoryear]{elsarticle}

\usepackage{amssymb}
\usepackage{lipsum}
\usepackage{amsmath}

\usepackage[utf8]{inputenc}

\usepackage{booktabs}
\usepackage{adjustbox}
\usepackage{tablefootnote}
\usepackage{comment}
\usepackage[mathscr]{euscript} %para tipografía especial

\usepackage[T1]{fontenc}
\usepackage{graphicx}   
\usepackage{caption}     
\usepackage{subcaption}

\usepackage{color}
\usepackage{multirow}
\usepackage{multicol}
\usepackage{graphicx}
\usepackage{hyperref}

\usepackage{float}
\hypersetup{
    colorlinks=true,
    linkcolor=blue,
    filecolor=blue,      
    urlcolor=blue,
    citecolor=blue,
}
\usepackage{txfonts}

\journal{High Energy Astrophysics}

\begin{document}

\begin{frontmatter}



\title{Effects of Bethe-Heitler pair production \\in ultraluminous X-ray sources}


\author[IAR,FCAG]{Gustavo E. Romero}
\ead{gustavo.esteban.romero@gmail.com}
\author[IAR,FCAG]{Lucas M. Pasquevich}
\author[IAR,FCAG]{Leandro Abaroa}

\affiliation[IAR]{organization={Instituto Argentino de Radioastronomía (CCT La Plata, CONICET; CICPBA; UNLP)},%Department and Organization
            addressline={Cno. Gral. Belgrano}, 
            city={Villa Elisa},
            postcode={1894}, 
            state={Buenos Aires},
            country={Argentina}}
            
\affiliation[FCAG]{organization={Facultad de Ciencias Astronómicas y Geofísicas, Universidad Nacional de La Plata},%Department and Organization
            addressline={Paseo del Bosque s/n}, 
            city={La Plata},
            postcode={B1900FWA}, 
            state={Buenos Aires},
            country={Argentina}}

\begin{abstract}
Some black holes in X-ray binaries accrete at rates far above the Eddington limit. In this supercritical regime, photons are trapped in a radiation-dominated, geometrically thick disk. The innermost regions form a complex environment of intense radiation, strong magnetic fields, and powerful outflows, where radiation-driven winds expel large amounts of mass. These conditions suppress primary relativistic electrons within the transparent funnel along the black hole’s spin axis. We show that high-energy electrons can instead arise as secondary pairs from Bethe-Heitler interactions between relativistic protons and ambient photons. Using self-similar models of accretion disks with strong winds of ultraluminous X-ray sources (ULXs), we compute particle acceleration via magnetic reconnection and diffusive shocks, evaluate energy losses, and assess the efficiency and spectral imprint of Bethe-Heitler pair production. Our results suggest that secondary pairs can yield nonthermal radiation in the 0.1-100 MeV range with luminosities from $10^{34}$ up to $10^{38}$ erg s$^{-1}$. This emission could be detectable by future MeV instruments from Galactic ULXs, offering evidence of relativistic protons in their inner funnels and revealing misaligned, otherwise hidden, super-Eddington sources in the Milky Way.

\end{abstract}




\end{frontmatter}

%\tableofcontents



\section{Introduction}
\label{sec:intro}

A fundamental parameter in the characterization of X-ray binary systems is the accretion rate of matter onto the black hole (BH). Depending on the ratio of the actual accretion rate to the Eddington rate, accretion proceeds in three distinct regimes.

At moderate accretion rates, the system enters the standard disk regime \citep{shakura1973}, characterized by an optically thick, geometrically thin accretion disk.

At very low accretion rates, the inner region of the disk inflates and becomes optically thin. In this case, the accreting gas is advected into the BH before it can cool efficiently, resulting in very high temperatures: approximately $T_{\rm e} \sim 10^9$ K for electrons and $T_{\rm p} \sim (m_{\rm p}/m_{\rm e}) , T_{\rm e} \sim 10^{12}$ K for protons. This regime, commonly referred to as an advection-dominated accretion flow (ADAF), is described by self-similar solutions developed by \citet{Ichimaru1977} and \citet{narayan1994}. The ADAF region resembles a hot, two-temperature corona surrounding the BH \citep[e.g.,][]{Bisnovatyi-Kogan1976,Liang-Price1977}. This plasma efficiently Comptonizes the soft radiation from the outer disk, producing hard X-ray emission extending up to photon energies of $\sim$150 keV, which is a hallmark of many X-ray binaries.

At the other extreme, when the accretion rate exceeds the Eddington limit, the disk becomes supercritical: geometrically thick and dominated by radiation pressure in its innermost regions. The luminosity saturates around the Eddington value, while the excess accreting matter is expelled in the form of a strong wind \citep[e.g.,][]{shakura1973,lipunova1999,fukue2004,ohsuga2005,fukue2010,Abaroaetal2024,Abaroa&Romero2024smbh}. Supercritical accretion is believed to power a variety of energetic astrophysical systems, including microquasars like SS-433, narrow-line Seyfert 1 galaxies, luminous quasars, transient episodes linked to tidal disruption events in supermassive BHs, and ultraluminous X-ray sources (ULXs) observed in external galaxies.

ULXs are believed to be accreting X-ray binaries hosting a compact stellar-mass object, such as a BH, accreting at super-Eddington rates (see \citealt{Fabrika_etal_2021_ULX} for a review). These systems exhibit X-ray fluxes which, if assumed to be isotropic, imply luminosities exceeding $L_{\rm X} > 10^{39}\rm{erg,s^{-1}}$ surpassing the Eddington limit for a BH with mass $M_{\rm BH} \lesssim 10 \; \rm{M}_\odot$. The corresponding Eddington luminosity is given by
\begin{equation}
L_{\rm{Edd}} = 1.26 \times 10^{38} \left( \frac{M_{\rm BH}}{\rm M_\odot} \right)\,\,\rm{erg\,s^{-1}}.
\end{equation}
In such conditions, the accretion rate, $\dot{M}_{\rm input}$, must exceed the Eddington rate, defined as $\dot{M}_{\rm Edd} = L_{\rm Edd} / (\eta_{\rm acc} c^2)$, where $\eta_{\rm acc} \sim 0.1$ is the accretion efficiency and $c$ is the speed of light.

The high observed luminosity may also result from geometric beaming \citep{King2010}: radiation from the inner disk is collimated by a dense, optically thick wind that surrounds the accretion flow, leaving only a narrow, transparent funnel aligned with the BH spin axis. As a result, only the innermost regions of the disk contribute to the observed emission, while radiation from the rest of the disk is absorbed and reprocessed by the wind.

However, most ULX spectra in the X-ray band exhibit a combination of a thermal blackbody component and a nonthermal power-law tail \citep[e.g.,][]{Combi_etal2024,Cruz-Sanchez_ulx2025}. The thermal emission is attributed to the hot inner regions of the accretion disk, while the power-law component is generally interpreted as arising from Comptonization of soft photons by relativistic electrons within the magnetocentrifugal funnel, which is confined by the dense wind. Two key open questions remain: how relativistic electrons are accelerated within this funnel, effectively acting as a coronal Comptonizing region in ULXs, and whether relativistic protons are also injected and accelerated in this environment.

%Inside the funnel, relativistic particles accelerate in internal shocks and interact with the matter, magnetic, and radiation fields. In particular, the interaction of protons with the disk radiation fields produces high-energy electron-positron pairs via the Bethe-Heitler mechanism. Secondary leptons produced by hadronic interactions can be more energetic than primary electrons, and we expect that their interaction with ambient fields can produce gamma emission on the order of MeV energies.

%In this regime, above a critical radius, the disk inflates by ram pressure. The limit of accretion rate of a black hole is the Eddington rate, so the matter above these limits is ejected in the form of powerful radiation-driven winds.

%Particles can be accelerated to relativistic energies by shocks in the magnetocentrifugal funnel that forms above the BH, around the spin axis of the disk. 
Both analytical models and numerical simulations have explored the generation of outflows in supercritical accretion disks \citep[e.g.,][]{meier1979,eggum1985,fukue2005,okuda2005,Sadowski2014,McKinney2015,sadowski2015,sotomayor2019,Kitakietal2021,Abaroaetal2023,Abaroaetal2024}. These studies show that both strongly magnetized \citep{McKinney2015} and nonmagnetized \citep{sadowski2015} disks are capable of launching mildly relativistic outflows in supercritical systems. Accretion of nondipolar magnetic fields and baryon entrainment from clumpy structures \citep{McKinney2009,romero2020} likely contribute to mass loading at the base of these outflows.

The intense radiation fields near the BH in ULXs are expected to severely limit particle acceleration, especially for relativistic primary electrons, which suffer strong radiative losses that impose a low-energy cutoff in their spectrum. In contrast, protons may be accelerated to significantly higher energies, although not necessarily high enough to efficiently produce photomesons. Nevertheless, Bethe-Heitler \citep{bethe1934stopping} interactions with ambient photons can generate abundant secondary electron-positron pairs. In some cases, these secondary pairs may carry more energy than the primary electrons, and their subsequent inverse Compton (IC) cooling can produce MeV gamma rays that may escape the system, offering a potential observational signature of relativistic protons in the funnel. Moreover, the number of secondary pairs could exceed that of primary electrons by several orders of magnitude, resulting in enhanced nonthermal luminosities in the gamma-ray band. The detectability of this emission depends sensitively on various factors, including the accretion rate and the relative intensity of different photon fields.

In this work, we investigate the role of Bethe-Heitler interactions in ULXs by studying the injection of nonthermal particle populations into the funnels formed by winds in supercritical accretion disks. Our primary objective is to identify observable signatures of proton acceleration in these systems.

The paper is organized as follows. In the next section, we present a self-similar model for supercritical accretion disks with winds and define two representative ULX configurations. We then inject populations of relativistic particles near the BH and estimate the associated energy losses and resulting spectra for both primary electrons and protons within the funnel. In Sect. \ref{sec:B-H}, we compute the production of secondary electron-positron pairs through Bethe-Heitler interactions and analyze their energy evolution. The resulting spectral energy distributions (SEDs) are presented in Sect. \ref{sec:SED}. In Sect. \ref{sec: opacity}, we examine photon absorption due to the dense wind, particularly at large inclination angles. A discussion of the implications of our results and their potential observational signatures is provided in Sect. \ref{sec:discussion}. We conclude in Sec. \ref{sec:conclusions} with a summary and outlook.

%--------------------------------------------------------------------
\section{Supercritical disks and winds}
\label{sec:disk-wind}

\subsection{General conditions of supercritical disks}
\label{qualitative}

When the accretion rate to the BH becomes supercritical, the thin disk approximation breaks down and the disk becomes thick near the compact object \citep[e.g.,][]{abramowicz1980,jaroszynski1980,paczynsky1980,abramowicz1988}. Photons produced well inside the disk are trapped in the flow, that is, the photon diffusion timescale exceeds the accretion timescale \citep[e.g.,][]{begelman1978,ohsuga2003,ohsuga2005,Kitakietal2021}. %Photon trapping occurs when the photon diffusion timescale ($t_{\mathrm{diff}} = 3\tau H/c)$) exceeds the accretion timescale ($t_{\rm accr} = -r/v_{r}$), where $\tau$ is the vertical Thomson scattering optical depth for a half disk of height $H$, $r$ is the radial coordinate measured from the BH, and $v_r$ is the radial velocity of the fluid. 
%Trapped photons are advected inward with the accretion flux and removed from the system by the BH. This process reduces the luminosity and energy conversion efficiency of the accretion disk. 

\begin{comment}
The luminosity can be expressed by two contributions, one from the photon trapping region and the other from the outer standard disk \citep{ohsuga2003}:
\begin{equation}
L \sim 2 \int_{r_{\mathrm{in}}} ^{r_{\mathrm{trap}}} 2\pi r Q_{\mathrm{vis}} \left(\frac{r}{r_{\mathrm{trap}}}\right)^{1/2}\mathrm{d}r + 2\int _{r_{\mathrm{trap}}} ^{\infty} 2\pi r Q_{\mathrm{vis}} \mathrm{d}r,
\end{equation}
where $Q_{\mathrm{vis}}$ is the viscous heating rate, $r_{\mathrm{in}}$ is the inner edge of the disk, and $r_{\mathrm{trap}}$ is the photon-trapping radius given by
\begin{equation}
r_{\mathrm{trap}} = \frac{3}{2}\frac{\dot{M}}{\dot{M}_{\mathrm{crit}}}\frac{H}{r}r_{\mathrm{g}}.
\end{equation}
This radius can be written as $r_{\mathrm{trap}} =  1.5\, \dot{m}h r_{\mathrm{g}}$, where $\dot{m}$ and $h$ are the normalized accretion rate and half-disk height, respectively. 
\end{comment}

The photon trapping effect can be parameterized by considering that the advection heating is a fraction of the viscous heating, $Q_{\mathrm{adv}} = Q_{\mathrm{vis}} - Q_{\mathrm{rad}} = f Q_{\mathrm{vis}}$, where $Q_{\mathrm{rad}}$ is the radiative cooling rate and $f$ is the advection parameter, constant along the disk \citep{narayan1994}. The critical radius of the accretion disk is defined as the distance from the BH where the radiation force overcomes gravity \citep[see][]{fukue2004}:
\begin{equation}
r_{\mathrm{crit}}=\frac{9\sqrt{3}\sigma_{\mathrm{T}}}{16\pi m_{\mathrm{p}}c}\dot{M}_{\mathrm{input}} \approx 40\dot{m}_{\rm input}r_{\mathrm{g}},
\label{eq:rcrit}
\end{equation}
where $\sigma_{\rm T}$ is the Thomson cross section, $\dot{m}_{\rm input}=\dot{M}_{\rm{input}}/\dot{M}_{\rm{Edd}}$ is the normalized accretion rate and $r_{\rm g}=GM_{\rm BH}/c^2$ the gravitational radius. If $\dot{M}_{\rm{input}} \gg \dot{M}_{\mathrm{Edd}}$, the critical radius can extend to the entire accretion disk, while for $\dot{M}_{\rm input} \gtrsim \dot{M}_{\mathrm{Edd}}$ there are two different regions in the disk: for $r < r_{\mathrm{crit}}$ the disk is in a supercritical state and for $r > r_{\mathrm{crit}}$ the disk is standard.

In the inner disk there is a strong mass loss of the system in the form of a wind driven by radiation pressure, and the accretion rate is controlled by the Eddington rate. The latter condition implies that the accretion rate for $r < r_{\mathrm{crit}}$ varies as $\dot{M}(r)\propto r$, %\begin{equation}
%\label{eq: critical accretion rate}
%\dot{M}(r) = \frac{16 \pi c m_{\mathrm{p}}}{9 %\sqrt{3} \sigma_{\mathrm{T}}}r,
%\end{equation}
while it is constant for $r > r_{\mathrm{crit}}$, $\dot{M}(r) = \dot{M}_{\rm input}$. %The general picture of a supercritical disk is shown in Fig. \ref{fig:ULX_cartoon}. 

Toroidal magnetic fields are expected to develop in the innermost disk of X-ray binaries. There, the overflowing stream from the Roche lobe expands in toroidal and radial directions as it feeds the outer disk. The disk shear should substantially enhance the toroidal component \citep{liska2018}. In the innermost region, the energy density of the magnetic field becomes important, so we distinguish an additional part of the accretion disk given by $ r_{\mathrm{in}} < r <r_{\rm mag}\, (< r_{\mathrm{crit}}) $ \citep[see][]{akizuki2006}, where $r_{\rm mag}$ is the magnetospheric radius and $r_{\rm in}=6r_{\rm g}$ is the innermost disk radius of a nonrotating BH. 

Each region is characterized by a typical temperature. The innermost part has temperatures of $\sim 10^7$ K, the critical region $\sim 10^6$ K, and the standard disk $\sim 10^5$ K. The typical temperature of the ejected wind is a fraction of the temperature of the critical disk. 

%%%%%%%%%%%%%%%%%%%%%%%%%%%%%%%%%%%%%%%%%%%%%%%%%%%%%%%%%%%%%%%%%%%%%%%%

\begin{figure}[h!]
    \centering   \includegraphics[width=0.9\linewidth]{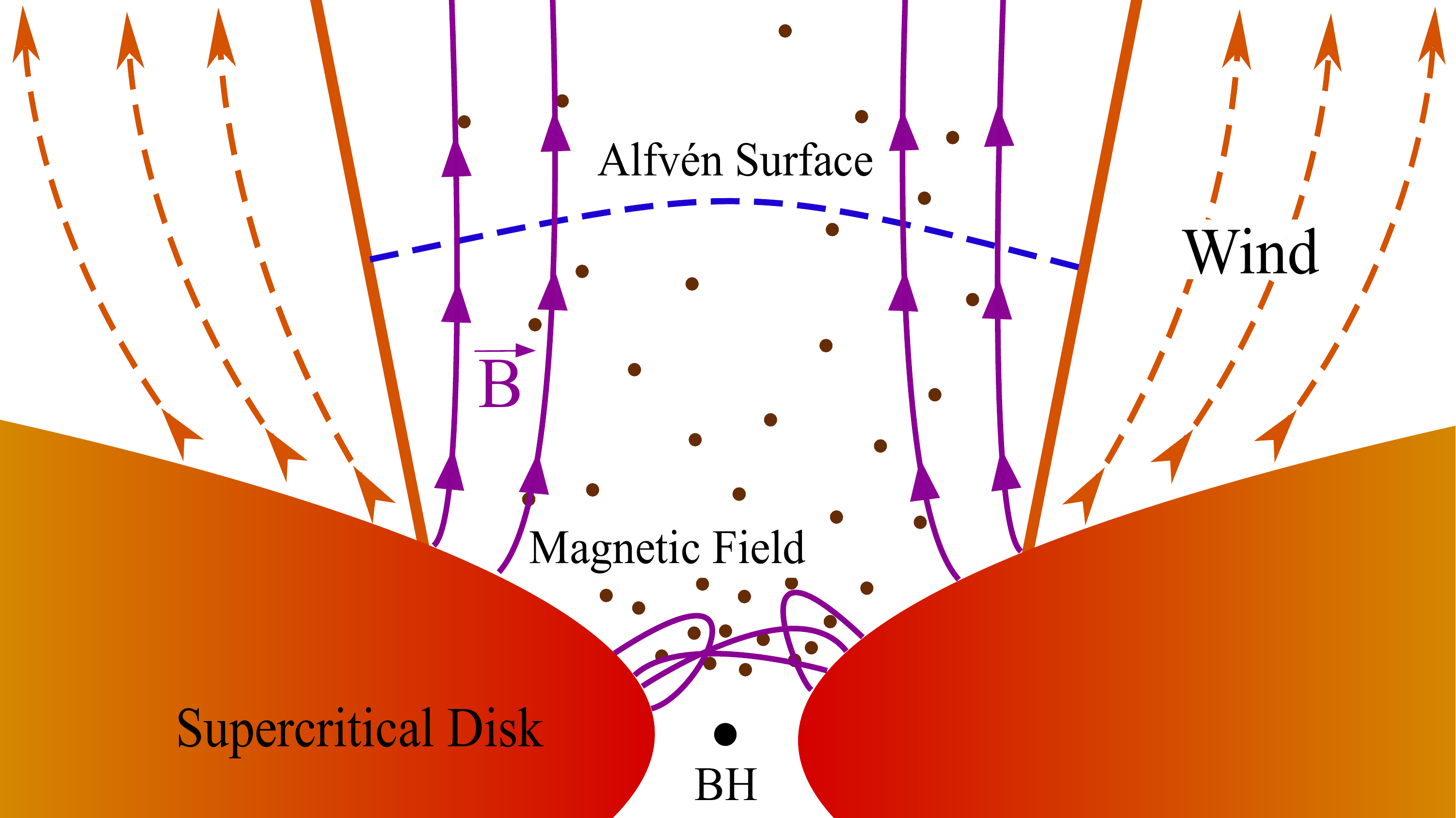}
    \caption{\small Diagram of the inner region of a ULX system. The dotted region is the funnel formed by the wind around the rotation axis of the BH. Not to scale.
}

    \label{fig:ULX_cartoon}
\end{figure}

%%%%%%%%%%%%%%%%%%%%%%%%%%%%%%%%%%%%%%%%%%%%%%%%%%%%%%%%%%%%%%%%%%%%%%%% 

\subsection{Disk model}
\label{sec: disk}

We assume a steady and axisymmetric disk where all physical quantities depend only on the distance to the BH on the equatorial plane, $r$. The continuity equation for the accreted gas is
\begin{equation}
\frac{1}{r}\frac{\mathrm{d}}{\mathrm{d}r}\left(r\Sigma v_{\rm r}\right) = 2 \dot{\rho}H,
\label{eq:continuity_disk}
\end{equation}
where $\Sigma=\Sigma_{0}r^{s}$ is the surface density ($\Sigma_0$ and $s$ are determined later), $v_{{\rm r}}$ the radial infall velocity, $H$ the scale height of the disk, and $\dot{\rho}$ the mass loss rate per unit volume.

The radial momentum equation can be written as
\begin{equation}
v_{{\rm r}}\frac{\mathrm{d}v_{{\rm r}}}{\mathrm{d}r} - \Omega^{2}r = -\Omega^{2}_{\mathrm{K}} - \frac{1}{\rho}\frac{\mathrm{d}}{\mathrm{d}r}\left(\rho c_{\mathrm{s}}^{2}\right),
\label{eq:radial_momentum_disk}
\end{equation}
where $\Omega$ is the angular velocity, $\Omega_{\mathrm{K}}$ the Keplerian angular velocity, and $c_{\mathrm{s}} \equiv \sqrt{p_{\mathrm{gas}}/\rho}$ the isothermal sound speed ($p_{\rm gas}$ is the gas pressure). The angular momentum equation is given by
\begin{equation}
v_{{\rm r}}\frac{\mathrm{d}}{\mathrm{d}r}\left(\Omega r^{2}\right) = \frac{1}{\rho r H} \frac{\mathrm{d}}{\mathrm{d}r}\left(\frac{\alpha \rho c_\mathrm{s}^{2}r^{3}H}{\Omega_{\mathrm{K}}}\frac{\mathrm{d}\Omega}{\mathrm{d}r}\right),
\label{eq:angular_momentum_disk}
\end{equation}
where we assumed there is no net radial or angular-momentum gain or loss associated with the wind \citep{fukue2004}. Here, $\alpha$ is the viscosity parameter.

The final two equations are those of hydrostatic balance in the vertical direction
\begin{equation}
\Omega_{\mathrm{K}}^{2}H^{2} = c_{\mathrm{s}}^{2},
\label{eq:balance_hydrostatic_disk}
\end{equation}
and the equation of energy:
\begin{equation}
\frac{\Sigma v_{{\rm r}}}{\gamma - 1}\frac{\mathrm{d}c_{\mathrm{s}}^{2}}{\mathrm{d}r} + 2Hc_{\mathrm{s}}^{2}\left(\dot{\rho} - v_{{\rm r}}\frac{\mathrm{d}\rho}{\mathrm{d}r}\right) = f \frac{\alpha \Sigma c_{\mathrm{s}}^{2}r^{2}}{\Omega_{\mathrm{K}}}\left(\frac{\mathrm{d}\Omega}{\mathrm{d}r}\right)^{2},
\label{eq:energy_disk}
\end{equation}
where $\gamma$ is the ratio of specific heats.

The effects of the magnetic field are neglected in all these equations. To include such effects, Eqs. (\ref{eq:radial_momentum_disk}), (\ref{eq:angular_momentum_disk}), and (\ref{eq:balance_hydrostatic_disk}) should be modified as in \cite{akizuki2006}. Additionally, we must introduce the induction equation. For a toroidal magnetic field, the equation is as follows:
\begin{equation}
v_{\rm r}\frac{\mathrm{d}c_{\mathrm{A}}^{2}}{\mathrm{d}r} + c_{\mathrm{A}}^{2}\frac{\mathrm{d}v_{{\rm r}}}{\mathrm{d}r} - \frac{c_{\mathrm{A}}^{2}v_{{\rm r}}}{r} = 2c_{\mathrm{A}}^{2}\frac{\dot{B_{\phi}}}{B_{\phi}} - c_{\mathrm{A}}^{2} \frac{2\dot{\rho}H}{\Sigma},
\label{eq:induction_disk}
\end{equation}
where $c_{\mathrm{A}} \equiv \sqrt{2p_{\mathrm{mag}}/\rho}$ is the Alfv\'en speed ($p_{\rm mag}$ is the magnetic pressure).

We solve the dynamic equations of the accreted fluid using the self-similar method of \cite{narayan1994}. The radial, azimuthal, sound, and Alfv\'en velocities for radii $r<r_{\rm crit}$ are given by:
\begin{equation}\label{eq:radial_vel}
v_{{\rm r}}(r) = -c_{1}\alpha \sqrt{\frac{GM_{\rm BH}}{r}},
\end{equation}
\begin{equation}
v_{\phi}(r) = c_{2} \sqrt{\frac{GM_{\rm BH}}{r}},
\end{equation}
\begin{equation}
c_{\mathrm{s}}^{2}(r) = c_{3}\frac{GM_{\rm BH}}{r},
\end{equation}
\begin{equation}
c_{\mathrm{A}}^{2}(r)  = 2\beta c_{3} \frac{GM_{\rm BH}}{r},
\end{equation}
where $\beta = p_{\mathrm{mag}}/p_{\mathrm{gas}}$ is the magnetization. 

The coefficients $c_{1}$, $c_{2}$, and $c_{3}$ are functions of the model parameters, $c_{i} = c_{i} (\alpha, \beta, f, s, \gamma)$. We refer to \cite{fukue2004} and \cite{akizuki2006} for the functional form of these coefficients.

Using the self-similar solutions, the mass-accretion rate for $r < r_{\mathrm{crit}}$ results
\begin{equation}
\label{eq: mass-accretion rate in critical disk}
\dot{M}(r) = -2\pi r \Sigma v_{{\rm r}} = \dot{M}_{\rm input}\left( \frac{r}{r_{\mathrm{crit}}}\right)^{s + 1/2}.
\end{equation}
The value of the exponent of the self-similar solutions in the critical disk model with mass loss in winds is $s = 1/2$. % (see Eqs. (\ref{eq: critical accretion rate}) and (\ref{eq: mass-accretion rate in critical disk})). 
The toroidal magnetic field is given by \citep{akizuki2006}:
\begin{equation}
    B_{\phi}=\sqrt{4\pi \Sigma_0 G M_{\rm BH} \frac
{\beta c_3}{\sqrt{(1+\beta)c_3}}}\,r^{s/2-1},
\end{equation}
where $\Sigma_0=\dot{M}_{\rm input}\,r_{\rm mag}(2\pi c_1 \alpha \,G^{1/2}\, M_{\rm BH}^{1/2}\,r_{\rm crit}^2)^{-1}$ (deduced from Eqs. (\ref{eq:radial_vel}) and (\ref{eq: mass-accretion rate in critical disk})).
Detailed formulae for computing the relevant physical parameters (thickness, temperature, radial velocity, and SED of the radiation emitted) can be found in \cite{fukue2004} and \cite{akizuki2006}.

We compute two different accretion disk scenarios characterized by the parameters given in Table \ref{table: 1} and accretion rates of $10\, \dot{M}_{\rm Edd}$ (scenario $A$) and  $10^3\, \dot{M}_{\rm Edd}$ (scenario $B$, hyperaccreting source; see e.g., \citealt{King2008}). We performed Monte Carlo simulations varying the parameters $\alpha$, $\beta$, $\gamma$, and $f$ within physically plausible ranges to obtain a set of parameters yielding predicted luminosities with a 95\% confidence level, thereby ensuring the robustness of our fiducial disk model.

The magnetospheric radius is estimated as in \cite{McKinney_etal_2012} (see also \citealt{narayan2003}):
\begin{equation}
\begin{array}{l}
    r_{\rm mag} = r_{\rm g} \left[12000 \left( \dfrac{3}{4} + \dfrac{n}{2} \right) \right]^{\frac{4}{3+2n}} \ \times \\ \\
    \times \ \ \epsilon_{-1}^{\frac{2}{3+2n}} m_8^{-\frac{6}{3+2n}} \dot{m}_{\rm H}^{-\frac{2}{3+2n}}\left( \dfrac{\Phi}{0.1\,{\rm pc^2\,G}} \right)^{\frac{4}{3+2n}},
\end{array}
\end{equation}
where $n$ stands for the relation between the accretion rate $\dot{M}(r)$ and the distance to the BH $r$ (in our case $n=1$, see Eq. (\ref{eq: mass-accretion rate in critical disk})), $\epsilon_{-1}=\epsilon/0.1$ ($\epsilon\sim0.01$ relates the advective to the free fall velocity of the plasma in the disk), $m_8=M_{\rm BH}/10^8M_{\odot}$, $\dot{m}_{\rm H}\approx0.15$ is the normalized accretion rate close to the BH (we take the value at the inner radius of the disk, $\dot{m}_{\rm H}=\dot{m}(r_{\rm in})$), and $\Phi$ is the magnetic flux near the BH (we assume a typical value for X-ray binaries of $\Phi=10^{-13}\,{\rm pc^2\,G}$, see e.g. \citealt{Justham_etal_2006}). Since the system is supercritical, $\dot{m}_{\rm H}$ is the same for any value of $\dot{M}_{\rm input}$. For the parameters assumed in our model, we obtain $r_{\rm mag}\approx 117r_{\rm g}$.

%%%%%%%%%%%%%%%%%%%%%%%%%%%%%%%%%%%%%%%%%%%%%%%%%%%%%%%%%%%%%%%%%%%%%%%%%%%%%%%%%%%%%
\begin{table}[h]
\centering                          % used for centering table
\caption{\small Parameters of the accretion disk model.}             % title of Table
\label{table: 1}      % is used to refer this table in the text

\begin{tabular}{l c}        % centered columns (4 columns)
\hline\hline                 % inserts double horizontal lines
Parameter & Value $(A, B)$  \\    % table heading 
\hline                        % inserts single horizontal line
   BH mass ($M_{\mathrm{BH}}$) & $10$  $M_{\odot}$ \\
 Eddington accretion rate  ($\dot{M}_{\mathrm{Edd}}$) & $2.2\times 10^{-7}$  $M_{\odot}\,\mathrm{yr}^{-1}$ \\
Eddington luminosity ($L_{\mathrm{Edd}}$) & $1.26\times 10^{39}$  $\mathrm{erg}\,\mathrm{s}^{-1}$ \\
Accretion rate ($\dot{M}_{\rm input}$) & $10, \;10^3$ $\dot{M}_{\rm Edd}$\\
   Gravitational radius  ($r_{\mathrm{g}}$) & $1.5\times 10^6$ $\mathrm{cm}$ \\
  Magnetized radius ($r_{\rm mag}$) & 117 $r_{\mathrm{g}}$\\
   Critical radius ($r_{\rm crit}$) & 400,  $4\times 10^4$ $r_{\mathrm{g}}$\\
Viscosity parameter ($\alpha$) & 0.01 \\  
 Advection parameter ($f$) &  0.5 \\
  Adiabatic index ($\gamma$) &  $4/3$ \\
   Self-similar exponent ($s$) & 1/2 \\   
   Magnetization of the inner disk ($\beta$) & 5\\
   \hline                                   %inserts single line
\end{tabular}
\end{table}
%%%%%%%%%%%%%%%%%%%%%%%%%%%%%%%%%%%%%%%%%%%%%%%%%%%%%%%%%%%%%%%%%%%%%%%%%%%%%%%

%%%%%%%%%%%%%%%%%%%%%%%%%%%%%%%%%%%%%%%%%%%%%%%%%%%%%%%%%%%%%%%%%%%%%%%%%%%%%%%%%

\begin{figure*}[h!]
\begin{multicols}{2}
\centering
    \includegraphics[width=8.2cm]{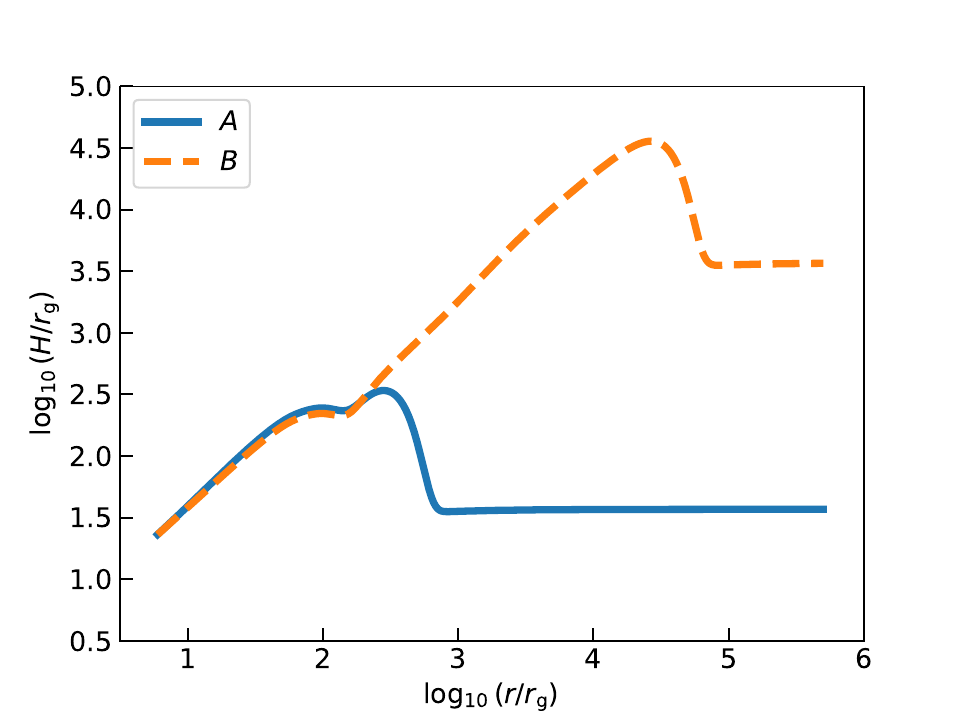}\par 
    \includegraphics[width=8.2cm]{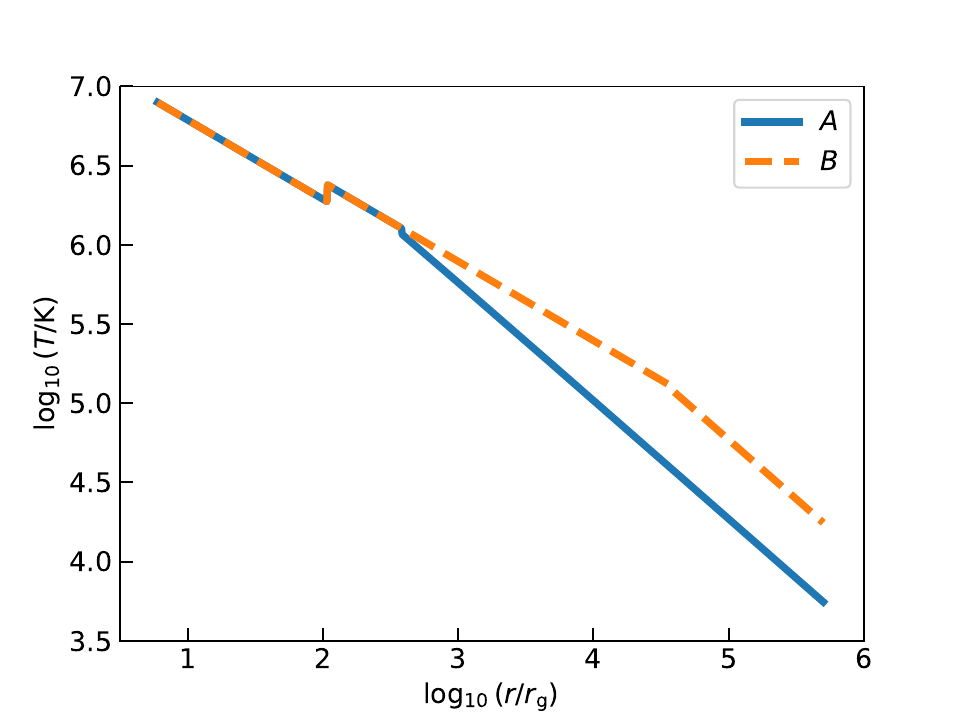}\par 
     \includegraphics[width=8.2cm]{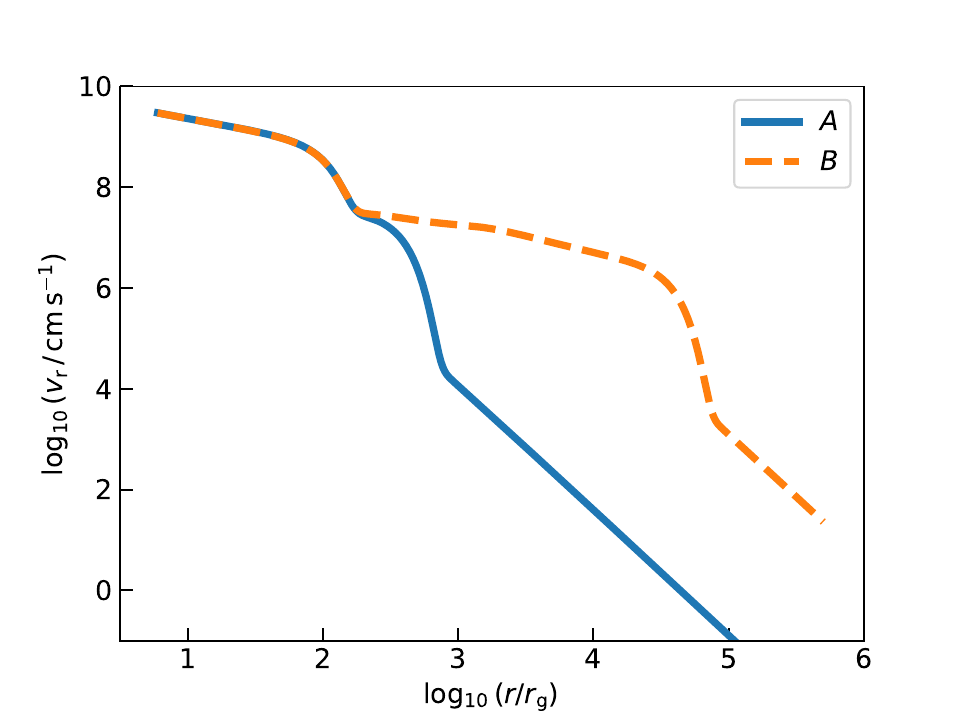}\par 
    \includegraphics[width=8.2cm]{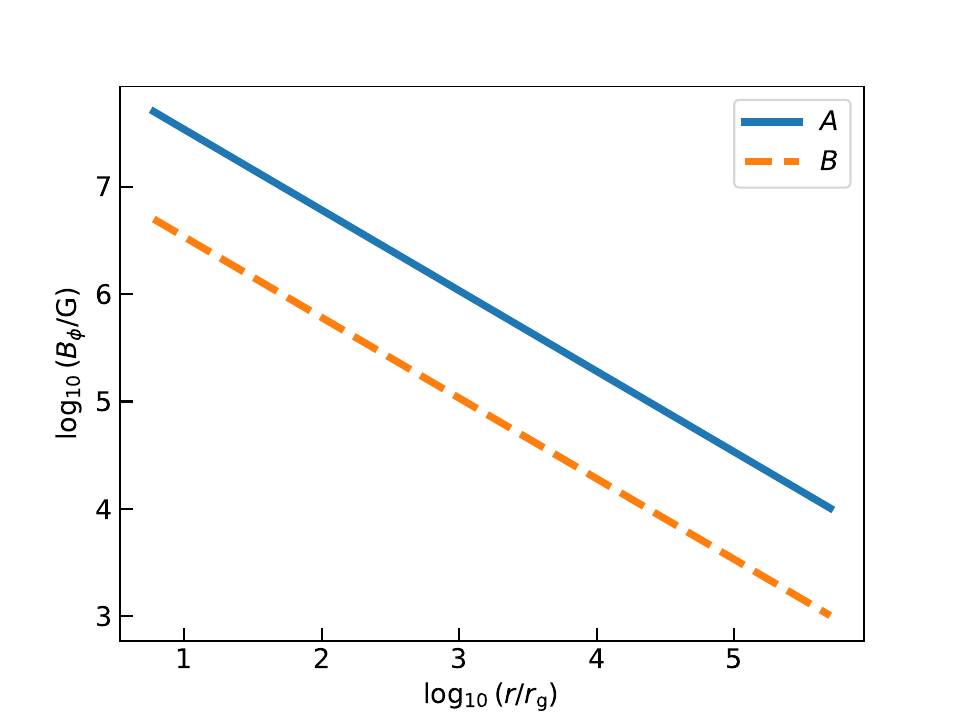}\par 
    \end{multicols}
\caption{Radial distribution of the thickness (top left), radial velocity (top right), temperature (bottom left), and toroidal magnetic field (bottom right) of the accretion disk. Thickness and radial distance to the BH are in units of gravitational radius. Radial velocity and magnetic field are in CGS units. The magnetospheric radius is at $\sim 117\,r_{\rm g}$, while the critical radius is at $\sim 400\,r_{\rm g}$ ($A$) and $\sim 40000\,r_{\rm g}$ ($B$).  We show results for scenarios $A$ ($\dot{m}_{\rm input}=10$, solid blue lines) and $B$ ($\dot{m}_{\rm input}=10^3$, dashed orange lines).}
\label{fig:properties_disk}
\end{figure*}
 
%%%%%%%%%%%%%%%%%%%%%%%%%%%%%%%%%%%%%%%%%%%%%%%%%%%%%%%%%
Figure \ref{fig:properties_disk} shows on a logarithmic scale the computed properties of the disk for scenarios $A$ (solid blue line) and $B$ (dashed orange line), as a function of $r$ in units of gravitational radius. The disk's thickness (top left) increases with $r$ up to a maximum at $r_{\rm crit}$, where it rapidly decays to the standard constant value. The radial velocity (top right) approaches a value of approximately $0.1c$ near the BH in both scenarios. Temperatures (bottom left) in the inner region can be as high as $T\sim10^7$ K, dropping below $T\sim10^5$ K beyond the critical radius. The magnetic field in the inner region can be very high close to the BH, with an order-of-magnitude difference between the two scenarios: $5\times10^7$ G ($A$) and $5\times10^6$ G ($B$). Note that the equations for the scale height and radial velocity are modified with softening factors at $r_{\rm mag}$ and $r_{\rm crit}$, where the disk's physical parameters change abruptly.

%%%%%%%%%%%%%%%%%%%%%%%%%%%%%%%%%%%%%%%%%%%%%%%%%%%%%%%%%%%%%%%%%%%%%%%%
% spectral energy distribution of the accretion disk
%%%%%%%%%%%%%%%%%%%%%%%%%%%%%%%%%%%%%%%%%%%%%%%%%%%%%%%%%%%%%%%%%%%%%%%%  

\begin{figure}[h!]
    \centering   \includegraphics[width=0.9\linewidth]{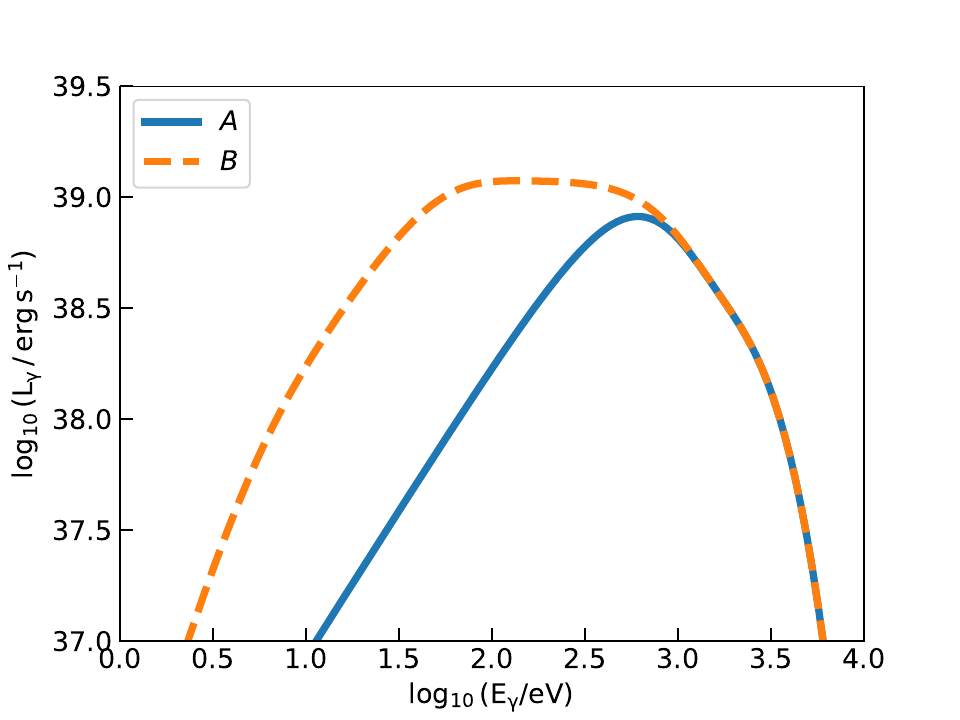}
    \caption{Unabsorbed and unbeamed thermal SEDs of the supercritical magnetized accretion disk for scenarios $A$ ($\dot{m}_{\rm input}=10$, solid blue line) and $B$ ($\dot{m}_{\rm input}=10^3$, dashed orange line). The maximum luminosity is limited to the Eddington luminosity, $\sim 10^{39}\,{\rm erg\,s^{-1}}$.}
    \label{fig:SED_accretion_disk}
\end{figure}

Figure \ref{fig:SED_accretion_disk} shows the SEDs of the disks without considering either the wind absorption or the geometrical beaming. In both scenarios, the intrinsic luminosity in the X-ray band is $\sim 10^{39}\,{\rm erg\,s^{-1}}$, that is, limited to the Eddington luminosity for the assumed BH mass. The disk SEDs, considering both absorption and beaming, are presented in Sect. \ref{sec:SED} along with all the radiative processes of the system. 

Unlike optically thin and very hot ADAFs, supercritical sources are not hot enough for efficient Comptonization. If Comptonization does occur, it must be produced by an additional component, such as a corona. However, the existence of a stable corona in super-Eddington sources is questionable, as radiation pressure blows away any plasma above the disk. The model of this ejected plasma is described below. %Nevertheless, we note that for smaller advection parameters it could be a hardening of the high-energy part of the spectrum. 

\subsection{Wind of the disk}
\label{wind}

We assume that the wind ejected in radii $r < r_{\mathrm{crit}}$ is highly ionized and blows off at a constant speed ($\beta_{\rm w} = v_{\rm w}/c$) and with a constant mass-outflow rate ($\dot{m}_\mathrm{w} = \dot{M}_{\mathrm{w}}/\dot{M}_{\mathrm{Edd}}$). 
\begin{comment}
The continuity equation for the wind is
\begin{equation}
\kappa_{\mathrm{co}}\rho_{\mathrm{co}}r_{\mathrm{g}} = \frac{\dot{m}_{\rm w}}{4\pi\Gamma_{\mathrm{w}}\, \beta_{\rm w}}\frac{r_{\mathrm{g}}^{2}}{R^{2}},
\label{eq:continuity_wind}
\end{equation}
where $\kappa_{\mathrm{co}} = \sigma_{\mathrm{T}}/m_{\mathrm{p}}$ is the opacity, $\rho_{\mathrm{co}}$ is the rest mass density, $\Gamma_{\mathrm{w}} = 1/\sqrt{1-\beta_{\rm w}^{2}}$ is the wind Lorentz factor, and $R=\sqrt{r^{2} + z^{2}}$ is the distance from the BH. 
\end{comment}
Since the disk self-regulates the luminosity approximately at the Eddington value, almost all the accreted mass is removed from the system through the wind (i.e., $\dot{m}_{\rm w}\approx \dot{m}_{\rm input}$). Its terminal speed can be estimated as $v_{\rm w}=c/\sqrt{40\, \dot{m}_{\rm input}}$ \citep{King2010}. We get velocities of $0.05c$ and $0.005c$ for scenarios $A$ and $B$, respectively. Therefore, $\beta_{\rm w}\ll1$ and $\gamma_{\rm w}\approx 1$, so the comoving properties of the wind are essentially equal to those observed. We assume that the wind is spherically symmetric, except for a narrow funnel around the BH axis, a common assumption for ULXs.

\begin{figure}[h!]
    \centering   \includegraphics[width=0.9\linewidth]{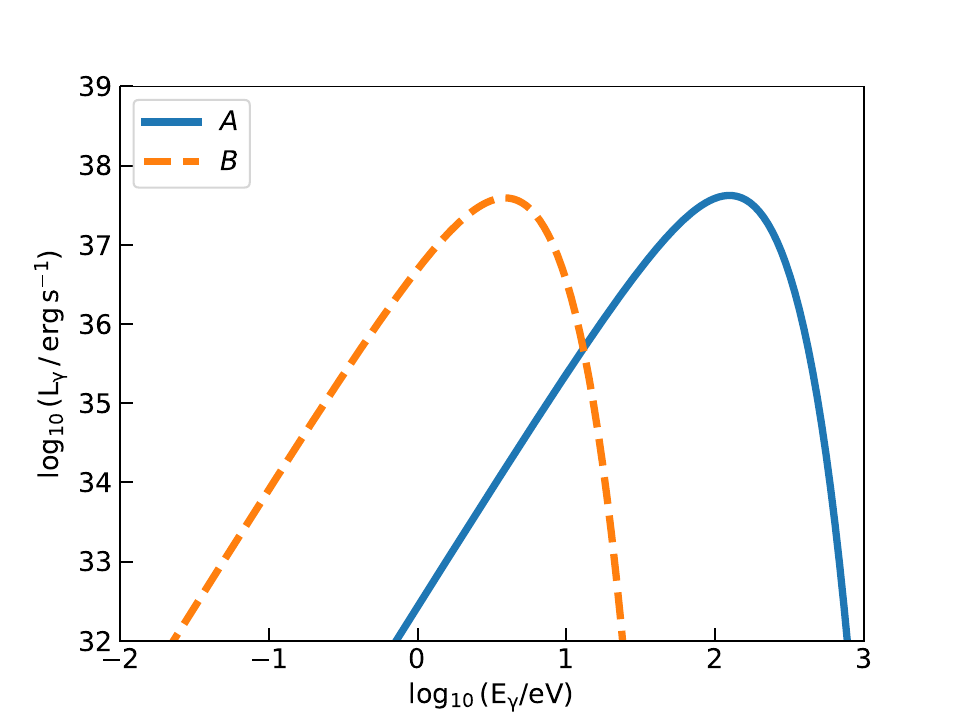}
    \caption{SEDs of the disk-driven wind for scenarios $A$ (solid blue line) and $B$ (dashed orange line). The luminosity in both cases is a fraction $\dot{e}\sim0.1$ of the Eddington luminosity.}
    \label{fig:wind}
\end{figure}

The wind mass-loss rate is much higher than the Eddington accretion rate onto the BH, so the wind is optically thick and emits blackbody radiation. The temperature $T_{\mathrm{w}}$ is given by \citep[see e.g.,][]{Abaroa-Romero2024RevMex}
\begin{equation}
\sigma_{\mathrm{T}}T_{\mathrm{w}}^{4} = \dot{e}\frac{L_{\mathrm{Edd}}}{4\pi R^{2}},
\label{eq:comoving_temperature_wind}
\end{equation}
where $\dot{e}\sim 0.1$ is the luminosity in Eddington units and $R=\sqrt{r^2+z^2}$.
\begin{comment}
The apparent photosphere of the wind is the surface where the optical depth $\tau$ becomes unity for an observer at infinity. The height of the apparent photosphere from the equatorial plane, $z_{\mathrm{ph}}$, can be calculated by solving the integral \citep[see e.g.,][]{Abaroa-Romero2024RevMex}
\begin{equation}
\tau_{\mathrm{ph}} = - \int	_{\infty} ^{z_{\mathrm{ph}}} \Gamma_{\mathrm{w}} \left(1 - \beta_{\rm w} \cos \theta \right)\kappa_{\mathrm{co}}\rho_{\mathrm{w, co}}\mathrm{d}z = 1.
\label{eq:height_photosphere_wind}
\end{equation}

Once the location of the apparent photosphere is determined, its temperature can be calculated from the comoving temperature by applying the appropriate boost:
\begin{equation}
T_{\mathrm{w,obs}} = \frac{1}{\Gamma_{\mathrm{w}}\,\left(1 - \beta_{\rm w} \cos \theta\right)}T_{\mathrm{w,co}},
\label{eq:observed_temperature_wind}
\end{equation}
where $\theta$ is the viewing angle measured from the $z$-axis. 
\end{comment}
Figure \ref{fig:wind} shows the SEDs of winds corresponding to cases $A$ and $B$. The maximum luminosities are $\lesssim  10^{38}\,{\rm erg\,s^{-1}}$. The wind photospheres are located at $\sim 2\times 10^3\,r_{\rm g}$ ($A$) and $\sim 2\times 10^6\,r_{\rm g}$ ($B$), with temperatures $\sim 10^5\,$K ($A$) and $\sim 10^4\,$K ($B$). 

We describe next the funnel formed by the wind walls along the spin axis of the BH and the particle gas flowing through it. 

%
%All matter ejected from the system at $r <r_{0}$ forms an outflow driven by radiation and magnetic pressure. In supercritical accretion disks the launch of jets by radiation pressure has been studied extensively \citep[see e.g.][]{piran1982,ohsuga2005,sadowski2015}. A mechanism of magneto centrifugal launching for a collimated jet might also operate if a poloidal component of the field develops in the innermost part of the disk. Magneto rotational instabilities can generate such poloidal component as demonstrated by \cite{liska2018}. 

%However, in what follows we will adopt a particle acceleration region inside the photosphere without discussing whether it is related or not to a jet.  \par

%%%%%%%%%%%%%%%%%%%%%%%%%%%%%%%%%%%%%%%%%%%%%%%%%%%%%%%%%%%%%%%%%%%%%%%%%%
% Properties of the wind disk
%%%%%%%%%%%%%%%%%%%%%%%%%%%%%%%%%%%%%%%%%%%%%%%%%%%%%%%%%%%%%%%%%%%%%%%%%%
%
%%%%%%%%%%%%%%%%%%%%%%%%%%%%%%%%%%%%%%%%%%%%%%%%%%%%%%%%%%%%%%%%%%%%%%%% 
%

\begin{comment}
%%%%%%%%%%%%%%%%%%%%%%%%%%%%%%%%%%%%%%%%%%%%%%%%%%%%%%%%%%%%%%%%%%%%%%%
% Table of outflows parameters
%%%%%%%%%%%%%%%%%%%%%%%%%%%%%%%%%%%%%%%%%%%%%%%%%%%%%%%%%%%%%%%%%%%%%%%
\begin{table*}[h!]
\caption{\small Parameters of the disk-driven wind (case of magnetized disk).}             % title of Table
\label{table_OutflowsParameters}      % is used to refer this table in the text
\centering                          % used for centering table
\begin{tabular}{l c c c}        % centered columns (4 columns)
\hline\hline                 % inserts double horizontal lines
Parameter & Symbol & Value ($A_1$, $A_2$) & Unit \\    % table heading 
\hline                        % inserts single horizontal line
   Total wind mass-loss rate & $\dot{M}_{\mathrm{wind}}$ & $2.2\times 10^{-6}$, $2.2\times 10^{-4}$ & $M_{\odot}\,\mathrm{yr}^{-1}$ \\     
   Total power of the wind & $L_{\mathrm{wind}}$ & $4.1\times 10^{42}$ & $\mathrm{erg\,s^{-1}}$ \\     
   Bolometric luminosity of the wind & $L_{\mathrm{wind}}$ & $4.1\times 10^{40}$ & $\mathrm{erg\,s^{-1}}$ \\     

      Wind velocity at $r_{\mathrm{cr}}$ & $v_{\mathrm{wind}} (r_{\mathrm{cr}})$ & $6\times 10^{3}$ & $\mathrm{km\,s^{-1}}$ \\
   Wind velocity at $r_{0}$ & $v_{\mathrm{wind}} (r_{0})$ & $4.2\times 10^{4}$ & $\mathrm{km\,s^{-1}}$ \\
\hline                                   %inserts single line
\end{tabular}
\end{table*}
\end{comment}

\subsection{Gas in the funnel}\label{subsec:funnel}

We assume that a gas of particles travels at semi-relativistic speeds through the transparent channel formed by the conical wind above the BH. These particles can enter the funnel directly from the innermost disk, where magnetic reconnection and accretion of non-dipolar magnetic fields can produce rapid flares that drag the particles along (see \citealt{romero2020}). 

In ULXs, the gas density must be low enough to ensure the funnel is optically thin -- an observer at infinity must be able to see the radiation from the innermost accretion disk, so the optical depth must satisfy $\tau<1$. The density of the gas in the funnel is given by $\rho_{\rm gas}=\dot{M}_{\rm gas}/\Omega\, z^2 \,v_{\rm gas}$, where $\dot{M}_{\rm gas}$ is the mass input to the funnel, $\Omega=2\pi(1-\cos{\vartheta})$ is the solid angle of the duct, and $z$ is the height above the BH. The semi-opening angle of the funnel $\vartheta$ is related, in principle, to the accretion mass-rate $\dot{m}_{\rm input}$ \citep{King2010}. We assumed semi-opening angles of $\vartheta_{A}=15^{\circ}$ (see e.g., \citealt{Veledina_etal_cygx3_2024NatAs}) and $\vartheta_{B}=5^{\circ}$ (the higher the mass loss in winds, the narrower the funnel), and the same gas velocity for both scenarios, $v_{\rm gas}=0.2 c$ \citep{Pinto_etal_2016Natur,Kosec_etal2018MNRAS,Fabrika_etal_2021_ULX,Abaroaetal2023}. 

We then solve the integral for the optical depth inside the funnel: 
\begin{comment}
\begin{multline}
     \tau=\int^\infty_{0} \gamma_{\rm gas}(1-\beta_{\rm gas} \cos{\Theta}) \, \kappa_{\rm co} \,\rho_{\rm co} {\rm d}z \approx \\
    \int^\infty_{0}  (1-\beta_{\rm gas} \cos{\Theta})\kappa_{\rm co} \,\rho_{\rm co}  {\rm d}z =1,
\end{multline}
\end{comment}
\begin{equation}
    \tau=\int^\infty_{0} \gamma_{\rm gas}(1-\beta_{\rm gas} \cos{\vartheta}) \, \kappa_{\rm co} \,\rho_{\rm co} {\rm d}z=1,
\end{equation}
where $\rho_{\rm co}\approx \rho_{\rm gas}$ and $\kappa_{\rm co}\approx\kappa=\sigma_{\rm T}/m_{\rm p}$ is the opacity, with $\sigma_{\rm T}$ the Thomson scattering and $m_{\rm p}$ the proton mass. We thus obtain the condition $\rho_{\rm gas}<5 \times 10^{-11}\,{\rm g\,cm^{-3}}$ with the parameters of our model for both scenarios. We assume this value for the gas density to calculate the radiative processes inside the funnel in two scenarios considered. We note that, since this is the maximum density allowed, the different results that we will obtain should be considered as upper limits.

\section{Relativistic primary particles}
\label{sec: primary particles}

We assume that there is a region of efficient particle acceleration inside the funnel and evaluate two possible mechanisms: diffusive shock acceleration (scenarios $A_1$ and $B_1$) and magnetic reconnection (scenarios $A_2$ and $B_2$). If shocks mediate the acceleration mechanism, then the region must be located at a distance such that the kinetic energy of the gas exceeds its magnetic energy; otherwise, the fluid would be mechanically incompressible. On the other hand, magnetic reconnection should occur near the BH, in order for the magnetic field to be sufficiently large.

Internal shocks caused by differential bulk motions occur  when the matter-energy density is greater than the magnetic energy density, $e_{\rm p}(z)=\dot{M}_{\rm gas}v_{\rm gas}/2\pi z^2 >B^2(z)/{8\pi}=e_{\rm m}(z)$. The strength of the magnetic field near the BH was calculated in Sect. \ref{sec: disk}. Its value for the two scenarios is $B(A_1)\sim5\times 10^{7}\,$G and $B(B_1)\sim 5\times 10^{6}\,$G, and decays as $\propto z^{-2}$ above the BH up to the Alf\'en radius. We find that $e_{\rm p}(z)>e_{\rm m}(z)$ at heights $z_{\rm acc}(A_1)\sim 200\,r_{\rm g}$ and $z_{\rm acc}(B_1)\sim 60\,r_{\rm g}$. We assume that the acceleration region is at $2000\,r_{\rm g}$ $(A_1)$ and $100\,r_{\rm g}$ $(B_1)$, where the magnetic fields are $B(A_1)\approx10^4\,{\rm G}$ and $B(B_1)\approx8\times 10^4\,{\rm G}$. The size of the acceleration region in both cases is $\Delta z_{\rm acc} =0.1\,z_{\rm acc}$.

%The magnetic field near the BH is calculated assuming equipartition between the magnetic and radiative energy densities, $e_{\rm m}(z)=e_{\rm r}=L/\Omega r^2 c$ (with $L$ the luminosity of the radiation field). The magnetic field in the funnel decreases with distance $z$ from the BH, $B(z)=B_0(z'/z)$, where $z'\approx 10\,r_{\rm g}$. On the other hand, the radiative energy density, $e_{\rm r}$, is calculated considering the thermal fields of the disk and the wind photosphere as blackbodies with temperatures of $T_{\rm disk}=10^7\,{\rm K}$ and $T_{\rm wind}=10^5\,{\rm K}$, respectively.

%We do not need to specify the exact mechanism for the injection of baryons in the funnel
The specific mechanism by which baryons are injected into the funnel remains uncertain, although several possibilities are discussed by, e.g., \cite{romero2020}. % We locate the region at a distance $z_{\rm acc}=1000\; r_{\rm g}$ above the inner disk. Its geometry is assumed to be a slab of radius $\mathbf{r_{\rm acc}=XXX}$ $r_{\rm g}$ and a thickness $\Delta z_{\rm acc}=100 r_{\rm g}<< z_{\rm acc}$. 
The physical conditions inside do not change significantly, so it can be considered as homogeneous. Basic parameters of this one-zone injector are given in Table \ref{table_acceleration}.

%%%%%%%%%%%%%%%%%%%%%%%%%%%%%%%%%%%%%%%%%%%%%%%%%%%%%%%%%%%%%%%%%%%%%%%
% Table of particle acceleration region 
%%%%%%%%%%%%%%%%%%%%%%%%%%%%%%%%%%%%%%%%%%%%%%%%%%%%%%%%%%%%%%%%%%%%%%%
\begin{table*}[h!]
\caption{\small Characteristic parameters of the particle acceleration region for models $A_1$, $B_1$, $A_2$ and $B_2$.}    % title of Table
\label{table_acceleration}      % is used to refer this table in the text
\centering                          % used for centering table
\begin{tabular}{l c c c c c c}        % centered columns (4 columns)
\hline\hline                 % inserts double horizontal lines
Parameter & Symbol & Value $(A_1)$ & Value $(B_1)$ & Value $(A_2)$ & Value $(B_2)$ &  Unit \\    % table heading 

\hline                        % inserts single horizontal line
   Height above the disk & $z_{\mathrm{acc}}$ & $2000$& $100$ & $10$ & $10$ & $r_{\rm g}$ \\     
   Radius & $r_{\rm acc}$ & $535$ & $9$ & $2.6$ & $0.8$ & $r_{\rm g}$ \\ 
   Size of acceleration region & $\Delta z_{\mathrm{acc}}$ &  $200$ &  $10$ &  $1$ &  $1$ & $r_{\rm g}$ \\     
   Thermal particle density & $n_{\mathrm{gas}}$ & $3.2\times 10^{12}$ & $1.3\times 10^{15}$ & $1.3\times 10^{16}$ & $1.3\times 10^{14}$ & $\mathrm{cm^{-3}}$ \\    
   Magnetic field & $B(z_{\mathrm{acc}})$ & $10^4$ & $8\times10^4$ & $5\times10^7$ & $5\times10^6$ & G \\ 
   Height of the wind photosphere & $H_{\rm ph}$ & 1900 & $2\times10^6$ & 1900 & $2\times10^6$ & $r_{\rm g}$\\
   Half-opening angle of the funnel & $\vartheta$ & 15 & 5 & 15 & 5 & degree\\
   Shock velocity & $v_{\rm s}$ & 0.3 & 0.3 & & & $c$\\
    Particle acceleration efficiency  & $\eta$ & $10^{-2}$ & $10^{-2}$ & 0.3 & 0.3   \\
    Fraction of power transferred to relativistic particles & $q_{\rm rel}$ & $0.1$ & $0.1$ & $0.1$ & $0.1$ \\
  Proton to electron ratio  & $a$ & $10^{3}$ & $10^{3}$ & $10^{3}$ & $10^{3}$  \\
  Index of the power law of injected particles & $p$ & 2 & 2 & 2 & 2  \\
\hline                                   %inserts single line
\end{tabular}
\end{table*}

%%%%%%%%%%%%%%%%%%%%%%%%%%%%%%%%%%%%
  
In the region $z_{\mathrm{acc}} \leq z \leq z_{\mathrm{max}}=z_{\mathrm{acc}} +\Delta z_{\rm acc}$ a small fraction of the total power $L_{\rm }$ (see Sect. \ref{subsec:funnel}) is transferred to relativistic particles:
\begin{equation}
L_{\mathrm{rel}} = q_{\mathrm{rel}}L_{\mathrm{}},
\label{equation_LuminosityRel}
\end{equation} 
where we adopt $q_{\mathrm{rel}} = 0.1$ and $L_{\mathrm{rel}}$ as the total power injected into both relativistic protons and electrons:
\begin{equation}
L_{\mathrm{rel}} = L_{\mathrm{p}} + L_{\mathrm{e}}.
\label{equation_LuminosityPart}
\end{equation} 
In addition, we assume that the mechanism is more efficient for protons, with a composition $a\sim m_{\rm p}/m_{\rm e}\approx 10^3$:
\begin{equation}
L_{\mathrm{p}} = aL_{\mathrm{e}}.
\label{equation_LuminosityDiv}
\end{equation} 
The kinetic power of the outflow, %\citep[see][]{drury1983}
\begin{equation}
L_{\rm gas} = \frac{\dot{M}_{\rm gas} v_{\rm gas}^2}{2},    
\end{equation}
is $10^{36}\,$erg s$^{-1}$ for scenario $A_1$ and $10^{35}\,$erg s$^{-1}$ for scenario $B_1$. On the other hand, the power carried by the magnetic field is \citep{2011ApJ...738..115D}: 
\begin{equation}
L_{\rm mag} = \frac{B^2 A(r) v_{\rm a}}{8\pi}.
\end{equation}
Here, $A(r) \approx 4\pi r_{\rm acc}^2$ is the surface area of the reconnection zone, and $v_{\rm a}=\sqrt{B^2/(4\pi m_{\rm p}n)}$ is the Alfvén speed ($\sim c$). The resulting magnetic power is $5 \times 10^{38}\,$erg s$^{-1}$ for scenario $A_2$ and $6 \times 10^{35}\,$erg s$^{-1}$ for scenario $B_2$.

The expected injection function is a power-law in the energy of the primary particles 
\begin{equation}
Q(E) = Q_{\mathrm{0}}{E^{-p}}, \qquad \left[Q\right] = \mathrm{erg}^{-1}\mathrm{s}^{-1}\mathrm{cm}^{-3},
\label{equation_InyectionParticles}
\end{equation}
where $p = 2$ for scenarios $A_1$ and $B_1$.

In the magnetically-dominated regime, the injection spectral index depends on the magnetization of the system ($\sigma_{\rm mag}$): % \citep{2015SSRv..191..545K}

\begin{equation}
    \sigma_{\rm mag}=\frac{B^2}{4\pi \,m \, n \, c^2 w_n}>1,
\end{equation}
where $B$ is the magnetic field, $n$ is the total particle number density, and $w_n = \gamma + P/(m \, n \, c^2)$ is the enthalpy per particle. For a cold gas $w_n \sim 1$, so $\sigma_{\rm mag} = B^2 / (4\pi\,n \,m  \, c^2)$.

For slightly magnetized systems ($\sigma_{\rm mag} \gtrsim 1$) the injection is $p \approx 4$; for moderate magnetization ($\sigma_{\rm mag} \approx 10$) the index is $p \approx 2$, while in highly magnetized systems, the spectral index can be as hard as $p \sim 1.5$ \citep{2015SSRv..191..545K}. With $\rho_{A_2} = 0.1\,\rho_{\mathrm{max},A_2}$ and $\rho_{B_2} = 0.001\,\rho_{\mathrm{max},B_2}$, we get $\sigma_{\rm mag} \approx 10$.

The normalization constant $Q_{\mathrm{0}}$ is obtained from
\begin{equation}
L_{(\mathrm{e,p})} = \int_{V} \mathrm{d}^{3}r \int ^{E_{(\mathrm{e,p})} ^{\mathrm{max}}} _{E_{(\mathrm{e,p})}^{\mathrm{min}}}\mathrm{d}E_{(\mathrm{e,p})}E_{(\mathrm{e,p})}Q_{(\mathrm{e,p})}(E_{(\mathrm{e,p})}),
\label{equation_LuminosityParticles}
\end{equation}
where $V$ is the volume of the acceleration region. The maximum energy $E^{\mathrm{max}}$ that a relativistic particle can attain results from balancing its acceleration rate with the cooling or escape rate.\par 

The acceleration rate for a charged particle in a magnetic field $B$ is \citep[see][]{drury1983,aharonian2004}:
\begin{equation}
t_{\mathrm{acc}}^{-1} = \frac{\eta e c B}{E},
\label{equation_AccelerationRate}
\end{equation}
where $E$ is the energy of the particle and $\eta$ is a parameter that characterizes the efficiency of the acceleration mechanism. In the case of diffusive shocks, we typically assume an efficiency of $10^{-2}$, while in the case of magnetic reconnection, the efficiency is \citep{drury1983, 2011ApJ...738..115D,2012A&A...542A...7V}:

\begin{equation}
    \eta_{\rm mag} \sim 0.1 \frac{r_{\rm g}c}{D(E)} \left( \frac{v_{\rm rec}}{c} \right)^2,
\end{equation}
where the diffusion coefficient in the Bohm regime is $D(E) = r_{\rm gy} c / 3$, and $r_{\rm gy} = E / (eB)$ is the gyroradius of the particle. In violent reconnection events, the reconnection speed approaches the Alfvén speed, i.e., $v_{\rm rec} \sim v_{\rm a}$. Since $v_{\rm a} \sim c$ in our scenario, the resulting acceleration efficiency is $\eta_{\rm mag} = 0.3$.

The cooling rate of the particles is equal to the sum of their radiative and adiabatic cooling rates:
\begin{equation}
t_{\mathrm{cool}}^{-1} = t_{\mathrm{rad}}^{-1} + t_{\mathrm{ad}}^{-1}.
\label{equation_CoolingRate}
\end{equation}

We consider electron losses due to synchrotron radiation, IC scattering, and relativistic Bremsstrahlung, and proton losses resulting from synchrotron radiation, inelastic proton-proton collisions ($pp$), and photon-proton interactions ($p\gamma$).
%\begin{equation}
%t_{\mathrm{rad}}^{-1} = t_{\mathrm{synchr}}^{-1} + t_{\mathrm{IC}}^{-1} + t_{\mathrm{Br}}^{-1},
%\end{equation}
%whereas for relativistic protons they are:
%\begin{equation}
%t_{\mathrm{rad}}^{-1} = t_{\mathrm{synchr}}^{-1} + t_{p\gamma, e^{\pm}}^{-1} + t_{p\gamma, \pi}^{-1} + t_{pp}^{-1}.
%\end{equation}
These losses are calculated using standard formulae given by \cite{blumenthal1970,begelman1990,kelner2006,kelner2008,romero2008}.

The adiabatic losses are caused by the expansion of the gas \citep{bosch-ramon2006},
\begin{equation}
t_{\mathrm{ad}}^{-1} = \frac{2}{3} \frac{v_{\mathrm{gas}}}{z}.
\label{equation_AdiabaticRate}
\end{equation}
The size of the acceleration region may be a constraint on the maximum energy if the particle gyroradius is greater than the linear size of such region (Hillas criterion).

In the "one-zone" approximation, the stationary particle distribution $N(E)$ can be obtained as the solution to the transport equation \citep[e.g.,][]{Ginzburg1964,Berezinskii1990}:
\begin{equation}
\frac{\partial }{\partial E}\left[\left.\frac{\mathrm{d}E}{\mathrm{d}t}\right\vert_{\mathrm{loss}}N(E)\right] + \frac{N(E,z)}{t_\mathrm{esc}} = Q(E),
\label{equation_ParticleDistribution}
\end{equation}
where $t_{\rm esc} \approx \Delta z_{\rm acc} / v_{\rm gas}$ is the particle escape time from the acceleration region in scenarios $A_1$ and $B_1$. In contrast, in scenarios $A_2$ and $B_2$, where particles are magnetically confined, the dominant escape mechanism is diffusion, characterized by the rate $t_{\rm diff}^{-1} = 2D(E)/r_{\rm acc}^2$, where $D(E)$ is the diffusion coefficient (we assume Bohm diffusion).

The exact solution to Eq. (\ref{equation_ParticleDistribution}) is given by \citep[see][]{khangulyan2007}:
\begin{equation}
N(E)= \left\vert\frac{\mathrm{d}E}{\mathrm{d}t}\right\vert_{\mathrm{loss}}^{-1}\int_{E}^{E^{\mathrm{max}}} \mathrm{d}E^{\prime}Q(E^{\prime})\exp\left(-\frac{\tau(E,E^{\prime})}{t_\mathrm{esc}}\right),
\label{equation_SolutionKhangaluyan}
\end{equation}
where
\begin{equation}
\tau(E,E^{\prime})=\int_{E}^{E^{\prime}} \mathrm{d}E^{\prime \prime} \left\vert\frac{\mathrm{d}E^{\prime \prime}}{\mathrm{d}t}\right\vert_{\mathrm{loss}}^{-1}.
\end{equation}

\begin{figure*}[h!]
    \centering
    % Fila 1: A1
    \begin{minipage}{0.45\textwidth}
        \centering
        \includegraphics[width=\linewidth]{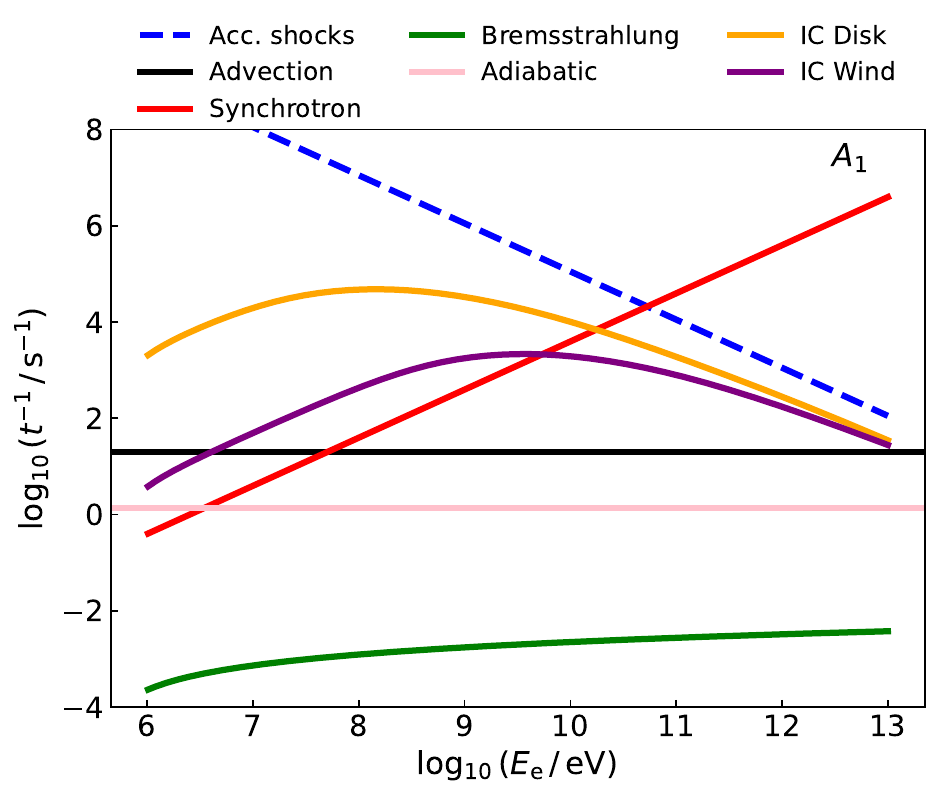}
        
    \end{minipage}
    \hfill
    \begin{minipage}{0.45\textwidth}
        \centering
        \includegraphics[width=\linewidth]{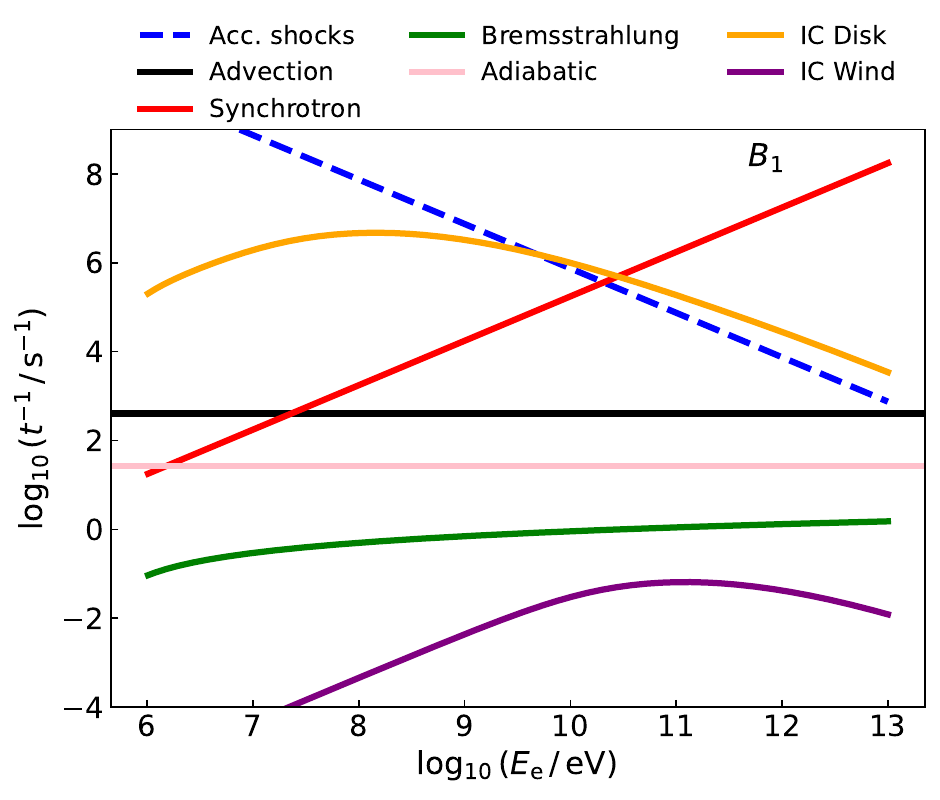}
        
    \end{minipage}

    \vspace{0.3cm} % Espacio entre filas

    % Fila 2: Protones
    \begin{minipage}{0.45\textwidth}
        \centering
        \includegraphics[width=\linewidth]{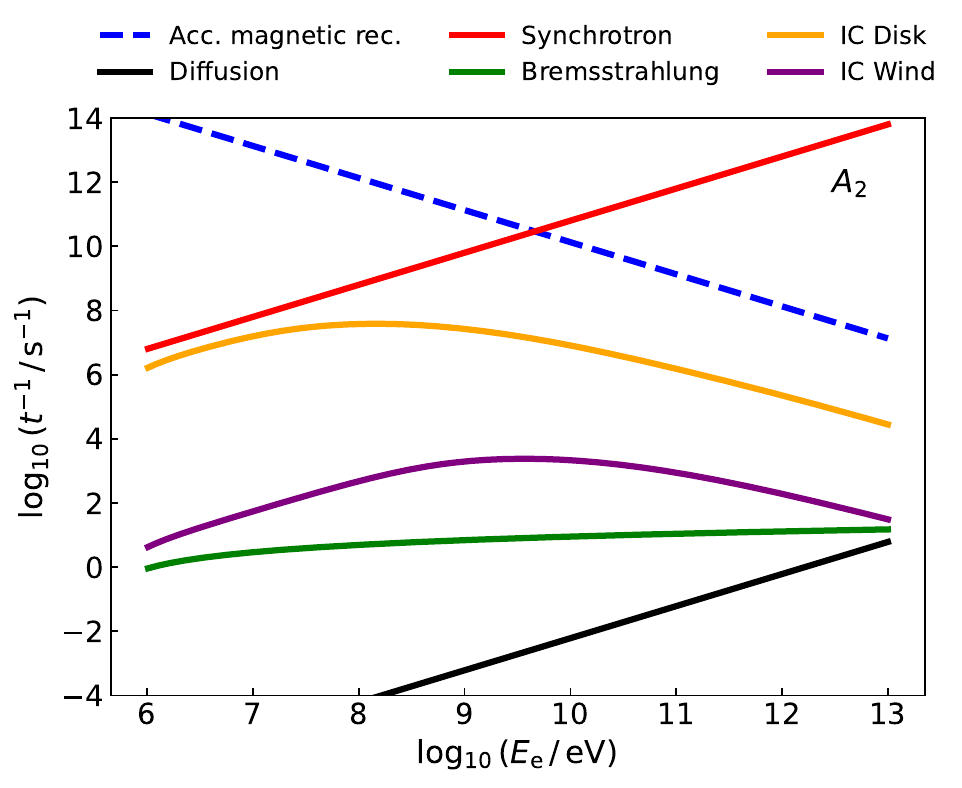}
       
    \end{minipage}
    \hfill
    \begin{minipage}{0.47\textwidth}
        \centering
        \includegraphics[width=\linewidth]{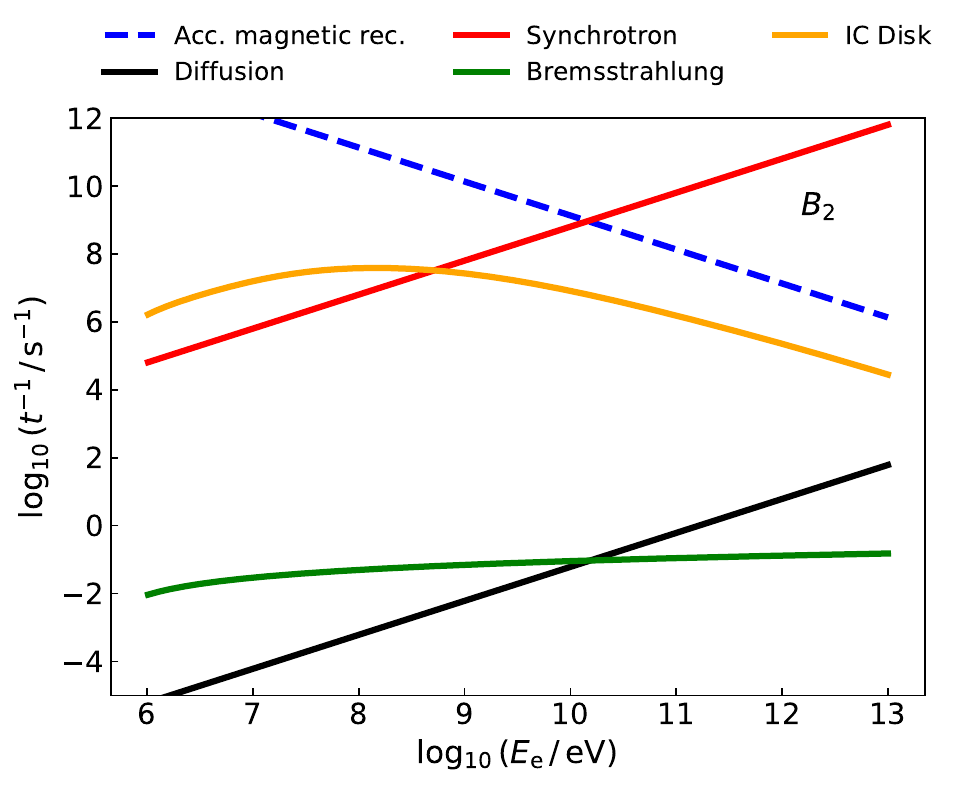}
        
    \end{minipage}

\caption{\small Timescales for the acceleration, cooling, and escape of relativistic electrons. 
\textbf{Top left:} Electron rates for $\dot{m}=10$ at $z=2000\,r_{\rm g}$ ($A_1$), where the acceleration mechanism is diffusive shock acceleration. The dominant cooling process is IC at low energies and synchrotron radiation at higher energies. The maximum electron energy is $\sim 50\,$GeV. 
\textbf{Top right:} Electron rates for $\dot{m}=1000$ at $z=200\,r_{\rm g}$ ($B_1$), with diffusive shocks as the acceleration mechanism. IC losses dominate across the entire energy range. The maximum energy is $\sim 6\,$GeV. 
\textbf{Bottom left:} Electron rates for $\dot{m}=10$ at $z=10\,r_{\rm g}$ ($A_2$). The acceleration mechanism is magnetic reconnection. Synchrotron radiation dominates the cooling, with a maximum energy of $\sim 3\,$GeV. 
\textbf{Bottom right:} Electron rates for $\dot{m}=1000$ at $z=200\,r_{\rm g}$ ($B_2$). The acceleration mechanism is magnetic reconnection. IC cooling dominates up to $\sim 1$ GeV, after which synchrotron losses dominate. The maximum energy reached is $\sim 20\,$GeV.}

    \label{fig:rates_electron}
\end{figure*}

Figure \ref{fig:rates_electron} shows the cooling and energy gains of primary electrons in the acceleration region for different scenarios. The maximum energies obtained are $\sim80\,{\rm GeV}$ ($A_1$), $\sim 5\,{\rm GeV}$ ($B_1$), $\sim 5\,{\rm GeV}$ ($A_2$), and $\sim 10\,{\rm GeV}$ ($B_2$). As expected, IC scattering of thermal photons from the disk and synchrotron radiation are the dominant losses. Bremsstrahlung, on the other hand, is negligible in all cases because of the low matter density inside the funnel.

\begin{figure*}[h!]
    \centering
    % Fila 1: A1
    
    \begin{minipage}{0.47\textwidth}
        \centering
        \includegraphics[width=\linewidth]{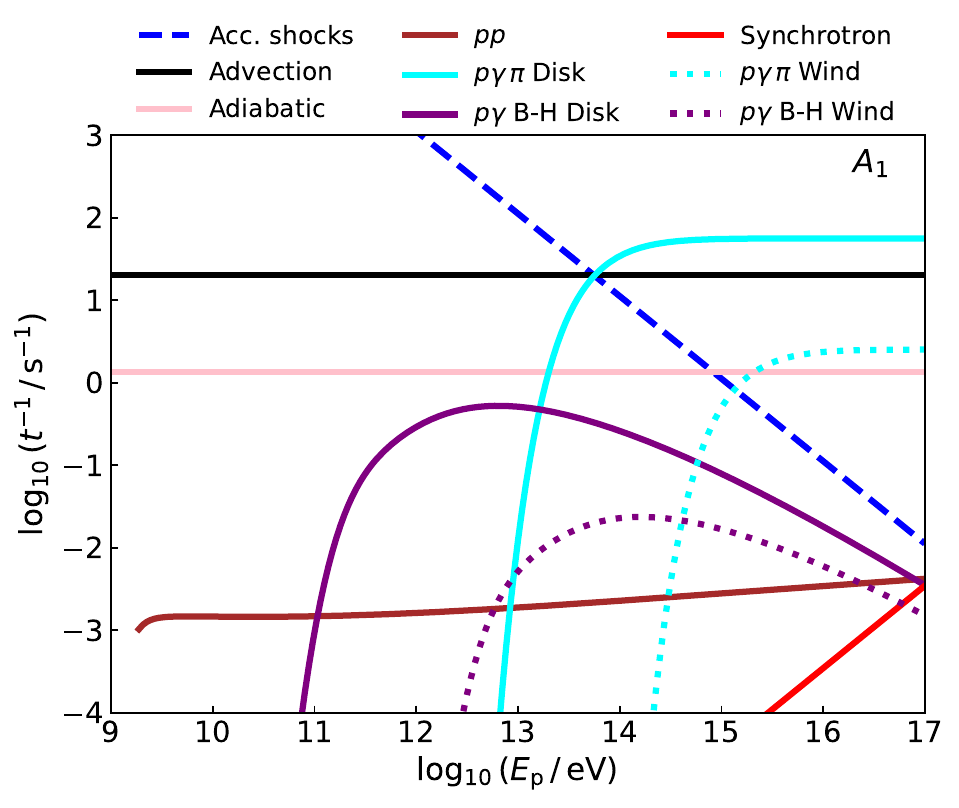}

    \end{minipage}
    \hspace{0.009\textwidth} % Espacio pequeño entre columnas
    \begin{minipage}{0.47\textwidth}
        \centering
        \includegraphics[width=\linewidth]{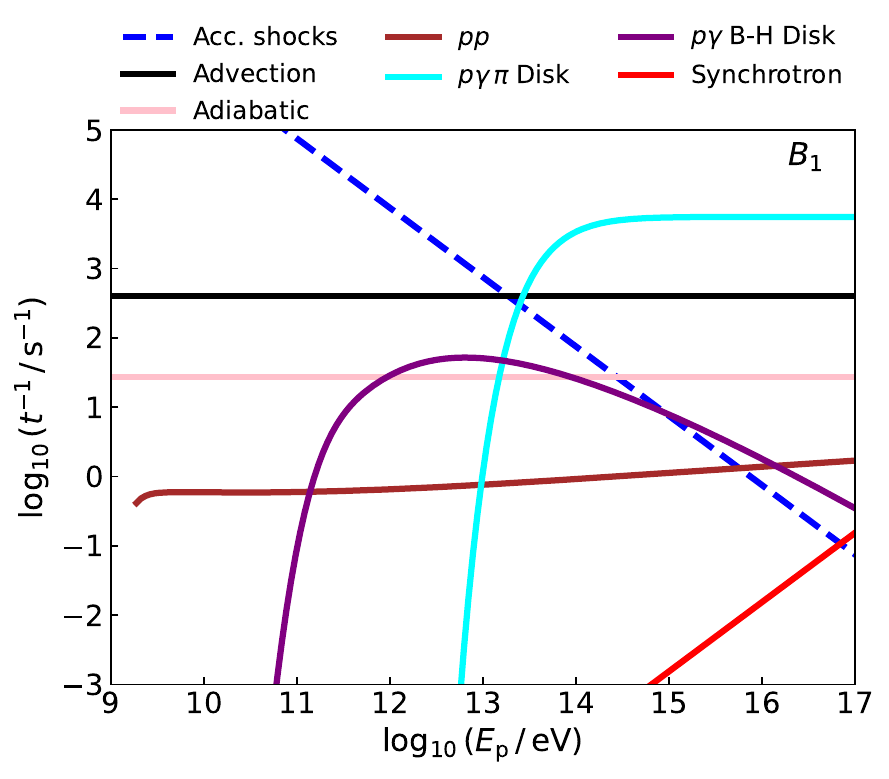}

    \end{minipage}

    \vspace{0.2cm} % Espacio entre filas

    % Fila 2: Protones
    \begin{minipage}{0.49\textwidth}
        \centering
        \hspace*{0.04\textwidth} % mueve la figura hacia la derecha
        \includegraphics[width=\linewidth]{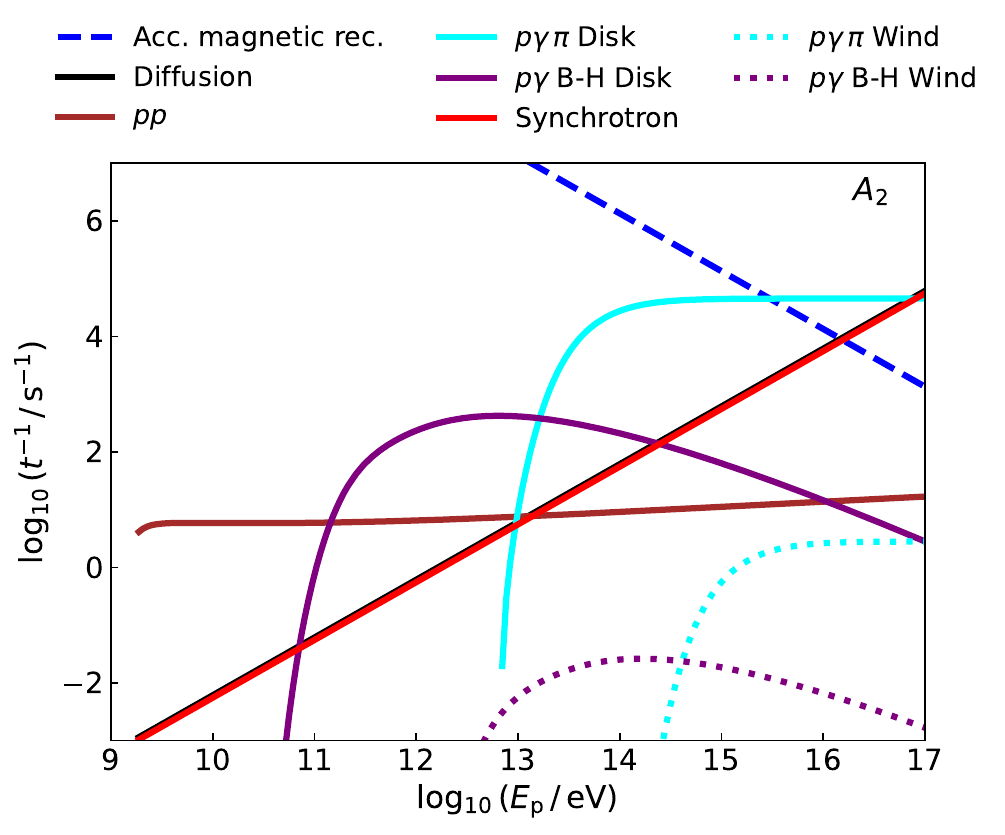}

    \end{minipage}
    \hfill
    \begin{minipage}{0.49\textwidth}
        \centering
        \includegraphics[width=\linewidth]{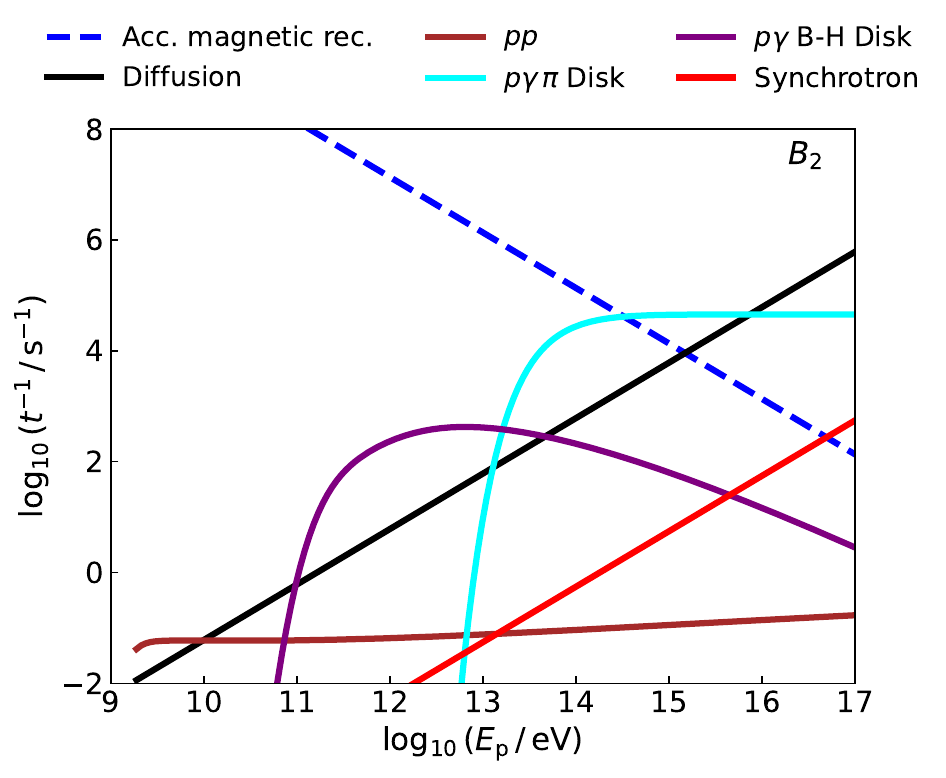}

    \end{minipage}

\caption{\small  Timescales for acceleration, cooling, and escape of relativistic protons. Scenarios as in Fig. \ref{fig:rates_electron}.
\textbf{Top left:} Convective escape dominates up to $\sim 20\,$TeV, beyond which photomeson production becomes the main energy loss channel. The maximum energy reached by protons is $\sim 50\,$TeV. 
\textbf{Top right:} Escape is the dominant process, and the maximum energy is $\sim 10\,$TeV. 
\textbf{Bottom left:} $pp$ interactions dominate at low energies, up to $\sim 100\,$GeV, followed by $p\gamma$ (photomeson) interactions at higher energies. The maximum energy is $\sim 5\,$PeV. 
\textbf{Bottom right:} At low energies, $pp$ interactions and diffusion dominate up to $\sim 100\,$GeV; at higher energies, photo-hadronic losses prevail. The maximum proton energy is $\sim 100\,$TeV.}
    \label{fig:rates_proton}
\end{figure*}

Figure \ref{fig:rates_proton} shows the cooling and energy gain rates for protons. Because of the intense radiation fields, the dominant radiative cooling mechanism is photo-hadronic interaction, as expected. In scenarios $A_1$ and $B_1$, particle escape is the dominant process. Photomeson production is only relevant in a narrow energy range near the maximum proton energy of $\sim 10\,{\rm TeV}$. In contrast, proximity to the disk in scenarios $A_2$ and $B_2$ results in stronger radiation and magnetic fields in the acceleration region. In these scenarios, radiative losses dominate, and photo-hadronic interactions are the primary cooling channel at energies above $\sim 100\,{\rm GeV}$. The maximum proton energies achieved are $\sim 5\,$PeV for $A_2$ and $\sim 200\,$TeV for $B_2$. Because of the low density, proton-proton losses are negligible in all scenarios. As shown in the figure, the production of electron-positron pairs via the Bethe-Heitler ($p\gamma$B-H) mechanism is significant in all cases. We will focus on this mechanism in the next section.

%%%%%%%%%%%%%%%%%%%%%%%%%%%%%%%%%%%%%%%%%%%%%%%%%%%%%
% Cooling and acceleration times for primary electrons and protons
%%%%%%%%%%%%%%%%%%%%%%%%%%%%%%%%%%%%%%%%%%%%%%%%%%%%%

%\begin{figure*}[h!]
% \centering
%    \subfloat{
%      \label{fig:Timessclaes}
%       \includegraphics[scale=0.53]{Paper Bethe-Heitler/figures/rates_pairs_10.pdf}}
%       \subfloat{
%      \label{fig:distribution_protons}
%      \includegraphics[scale=0.53]{Paper Bethe-Heitler/figures/rates_pairs_1000.pdf}}
%     \caption{\small Acceleration, cooling, and escape timescales for relativistic secondary electrons in the funnel. %The left plot is for $\dot{m}$=10, and the right plot is for $\dot{m}$=1000.}
%      \label{fig:rates_pairs}
%\end{figure*}

   %%%%%%%%%%%%%%%%%%%%%%%%%%%%%%%%%%%%%%%%%%%%%%%%%%%%%%%%%%%%%%%%%%%%%%%%  
% Distribution of primary relativistic particles.
%%%%%%%%%%%%%%%%%%%%%%%%%%%%%%%%%%%%%%%%%%%%%%%%%%%%%%%%%%%%%%%%%%%%%%%%
\begin{figure*}[h!]
 \centering
    \subfloat{
      \label{fig:distribution_electrons}
       \includegraphics[scale=0.5]{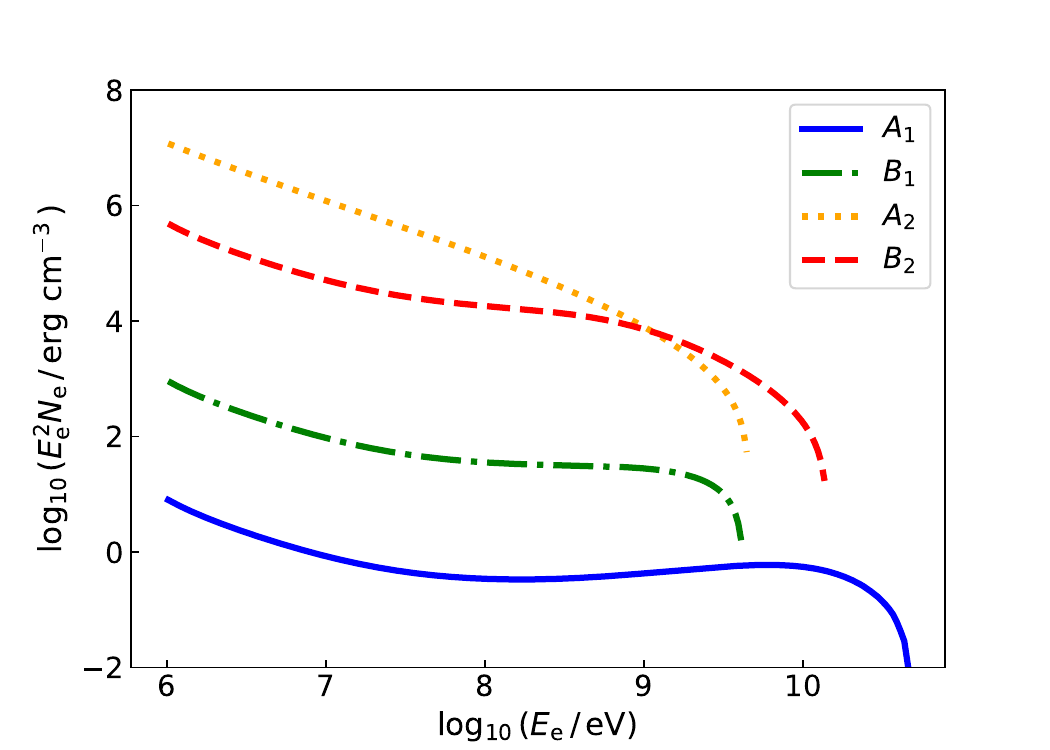}}
       \quad % Adds horizontal spacing between subfigures
       \subfloat{
      \label{fig:distribution_protons}
      \includegraphics[scale=0.5]{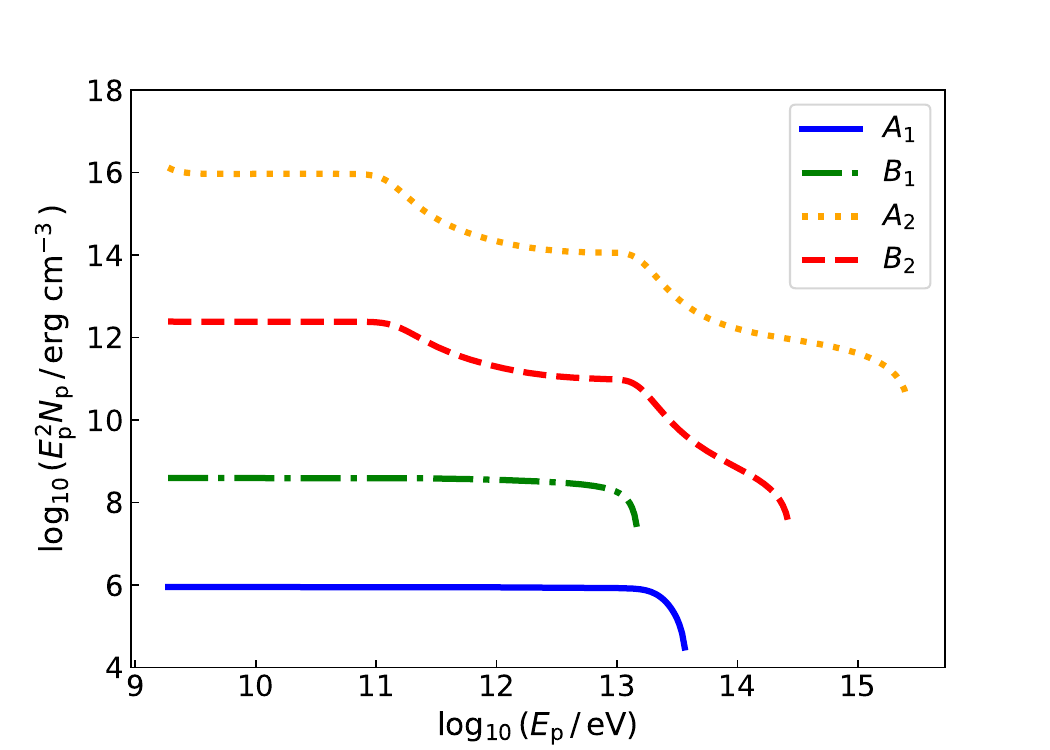}}
     \caption{\small Distribution of primary relativistic particles for the various models.
    \textbf{Left:} Distribution of primary electrons.
    \textbf{Right:} Distribution of protons.
 }
      \label{fig:Distribution_primary_particles}
\end{figure*}

   %%%%%%%%%%%%%%%%%%%%%%%%%%%%%%%%%%%%%%%%%%%%%%%%%%%%%%%%%%%%%%%%%%%%%%%%  
% Distribution of secondary relativistic particles.
%%%%%%%%%%%%%%%%%%%%%%%%%%%%%%%%%%%%%%%%%%%%%%%%%%%%%%%%%%%%%%%%%%%%%%%%
\begin{figure*}[h!]
    \centering 
    \includegraphics[scale=0.47]{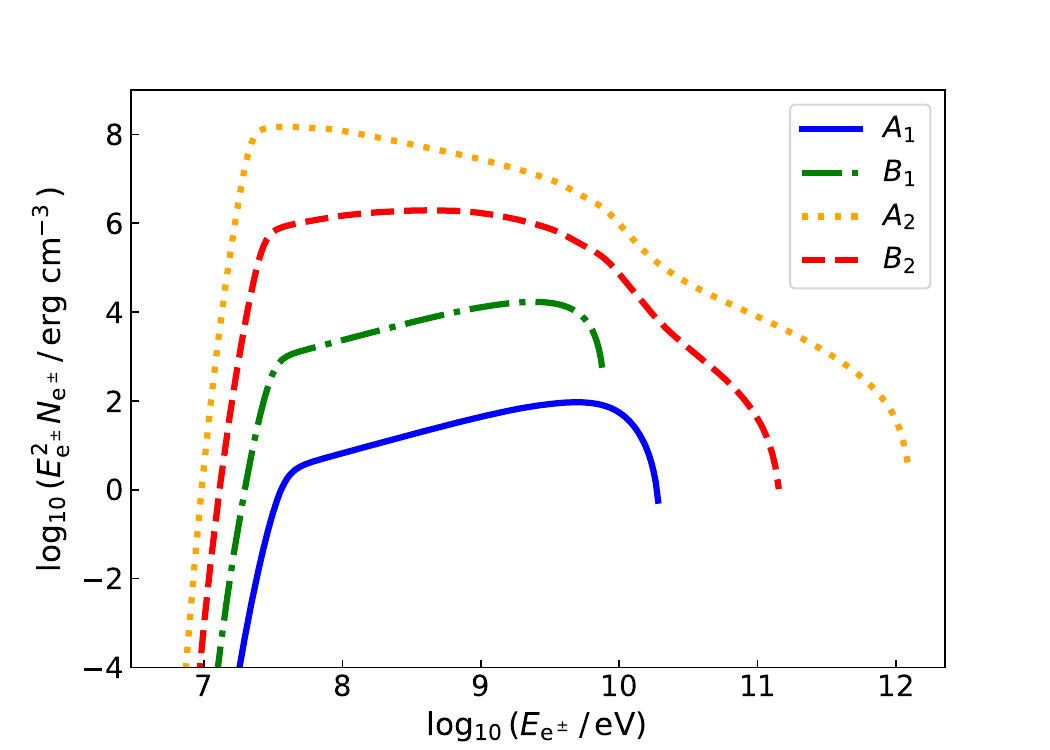}
    \caption{\small  Distribution of secondary $e^{\pm}$pairs produced via the Bethe-Heitler mechanism for all scenarios.
}
\label{fig:distribution_pairs}
\end{figure*}

\section{The Bethe-Heitler secondary pair injection}
\label{sec:B-H}

Primary electrons cool via synchrotron and IC radiation. The main interaction process for protons in the funnel involves proton-photon ($p\gamma$) collisions. This interaction has two main channels. The first is the Bethe-Heitler effect, also known as photo-pair production
\begin{equation}
p+\gamma\rightarrow p+e^-+e^+,
	\label{photopair}
\end{equation}  
with a threshold energy of $\sim1$ MeV for the photon in the proton rest frame. 

The second channel is photomeson production, which proceeds through two main branches: 
\begin{equation}
p+\gamma\rightarrow p+a\pi^0+b\left(\pi^++\pi^-\right),
	\label{photomeson1}
\end{equation}  
\noindent and
\begin{equation}
p+\gamma\rightarrow n+\pi^++a\pi^0+b\left(\pi^++\pi^-\right),
	\label{photomeson2}
\end{equation}  
\noindent with approximately the same cross-section and a threshold energy of $\sim145$ MeV. Here, $a$ and $b$ are the pion multiplicities. Charged pions decay in leptons and neutrinos,
\begin{equation}
\pi^+\rightarrow\mu^++\nu_\mu, \quad \mu^+\rightarrow e^++\nu_{\rm{e}}+\overline{\nu}_\mu
	\label{piondecay1}	
\end{equation}
\begin{equation}
\pi^-\rightarrow\mu^-+\overline{\nu}_\mu, \quad \mu^-\rightarrow e^-+\overline{\nu}_{\rm{e}}+\nu_\mu,
	\label{piondecay2}	
\end{equation}
whereas neutral pion decay yields photons: $\pi^0\rightarrow2\gamma$.

In this paper, we are mainly interested in the secondary pairs injected via the Bethe-Heitler mechanism. This process has been studied, for example, by \cite{chodorowski1992}, \cite{mucke_etal2000}, and \cite{mastichiadis2005}. The emissivity of pairs in the $\delta$-functional approximation is calculated as in \cite{romero2008}. The inelasticity can be approximated by its value at the threshold, $K_{p\gamma,\,e^\pm}=2m_{\rm e}/m_{\rm p}$. Therefore we have,  
\begin{align}
Q_{\rm e^{\pm}}\left(E_{\rm e^{\pm}}\right) 
&= 2 \int \mathrm{d}E_{\rm p} \, N_{\rm p}\left(E_{\rm p}\right)\,\omega_{\rm p\gamma,\,e^\pm}\left(E_{\rm p}\right)\, 
\delta\left(E_{\rm e^{\pm}} - \frac{m_{\rm e}}{m_{\rm p}} E_{\rm p} \right) \nonumber \\
&= 2 \frac{m_{\rm p}}{m_{\rm e}}\, N_{\rm p}\left( \frac{m_{\rm p}}{m_{\rm e}} E_{\rm e^{\pm}} \right)\, 
\omega_{\rm p\gamma,\,e^\pm} \left( \frac{m_{\rm p}}{m_{\rm e}} E_{\rm e^{\pm}} \right).
\label{emispairs}
\end{align}

\noindent An useful parametrization of the cross section $\sigma_{\rm p\gamma,\,e^\pm}$ can be found in \cite{Maximon1968}. Since the outflow is not relativistic $(\gamma_{\rm gas}\approx 1)$, the above expressions are the same for both the comoving and observer reference frames.  We calculate the emissivity of electron-positron pairs injected into the same region as the primary particles, using the same ambient fields.

\section{Particle distribution and SEDs}
\label{sec:SED}

We solve the transport equation to obtain the distribution of primary and secondary relativistic particles. The results for primary particles are shown in Figure \ref{fig:Distribution_primary_particles}, where we plot the distribution of primary electrons (left) and protons (right), for all scenarios. The distribution of secondary pairs produced via Bethe-Heitler mechanism is shown in Fig. \ref{fig:distribution_pairs}. The number of relativistic secondary pairs exceeds that of primary electrons by several orders of magnitude. The maximum energies of the Bethe-Heitler pairs are approximately 20 GeV ($A_1$), 9 GeV ($B_1$), 1 TeV ($A_2$), and 200 GeV ($B_2$).

The next step is to calculate the SEDs of the various nonthermal processes. We refer to \cite{romero2008} and references therein for the formulae used. Using the results presented in sections \ref{sec: primary particles} and \ref{sec:B-H}, we calculate the synchrotron and IC radiation for the primary electrons and secondary pairs, and the emission due to neutral pion decay produced by $p\gamma$ interactions. 

\begin{figure*}[h!]
 \centering
    \subfloat{
       \includegraphics[scale=0.4]{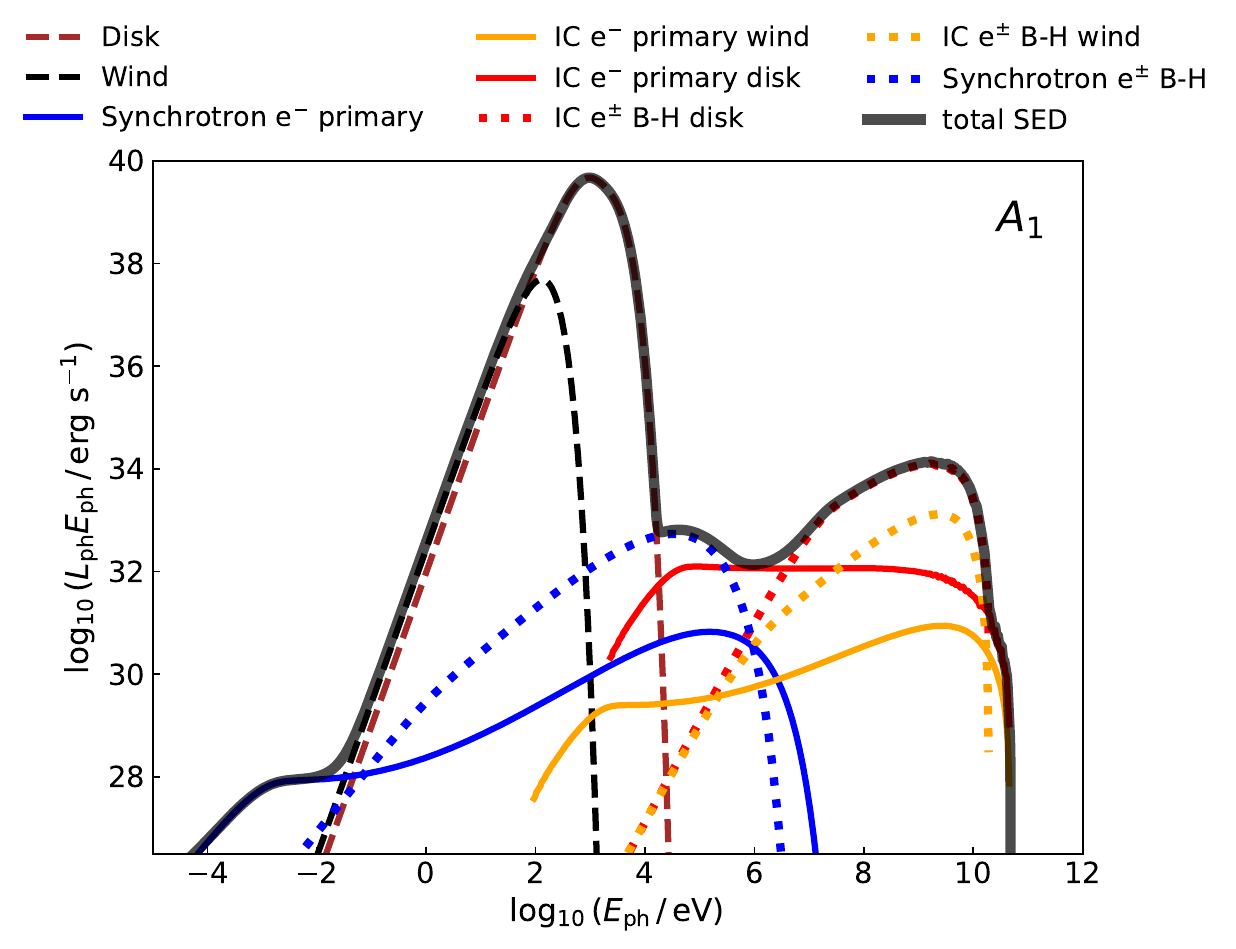}}
       \subfloat{
      \includegraphics[scale=0.4]{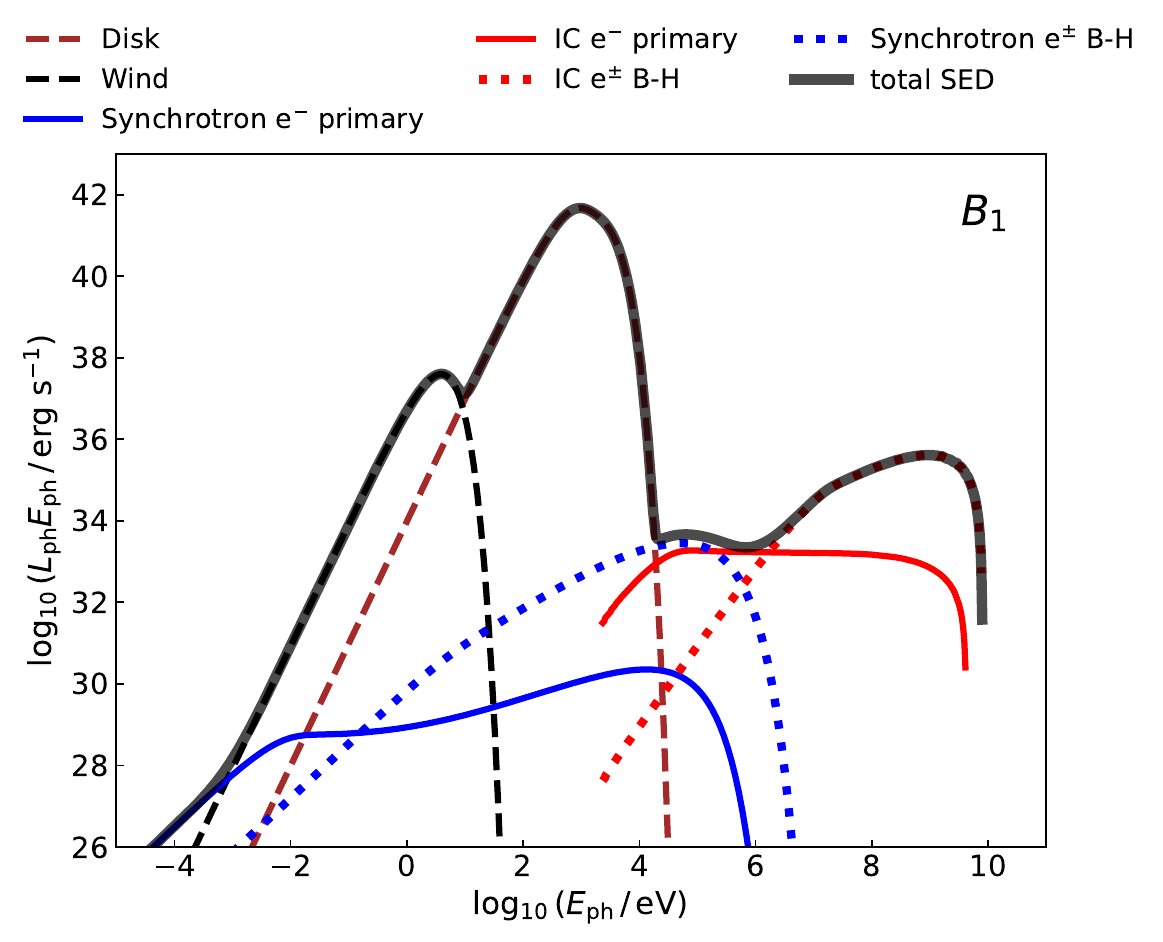}}
\caption{\small SEDs of the entire system for both accretion rates in the scenario of particle acceleration by diffusive shocks. Absorption effects are significant only above $100\,$GeV. The solid black line is the overall emission.}
      \label{fig:seds}
\end{figure*}

Figure \ref{fig:seds} shows the results for scenarios $A_1$ (left) and $B_1$ (right). Synchrotron and IC radiation from the secondary pairs are the dominant nonthermal processes, with maximum luminosities of $\sim 10^{34}\,{\rm erg\,s^{-1}}$ in scenario $A_1$ and $\sim 10^{36}\,{\rm erg\,s^{-1}}$ in scenario $B_1$, around $\sim1\,$GeV. This emission dominates at energies of $\sim 1\,\mathrm{MeV}$, closely followed by IC radiation by primary electrons.
 
\begin{figure*}[h!]
 \centering
    \subfloat{
       \includegraphics[scale=0.4]{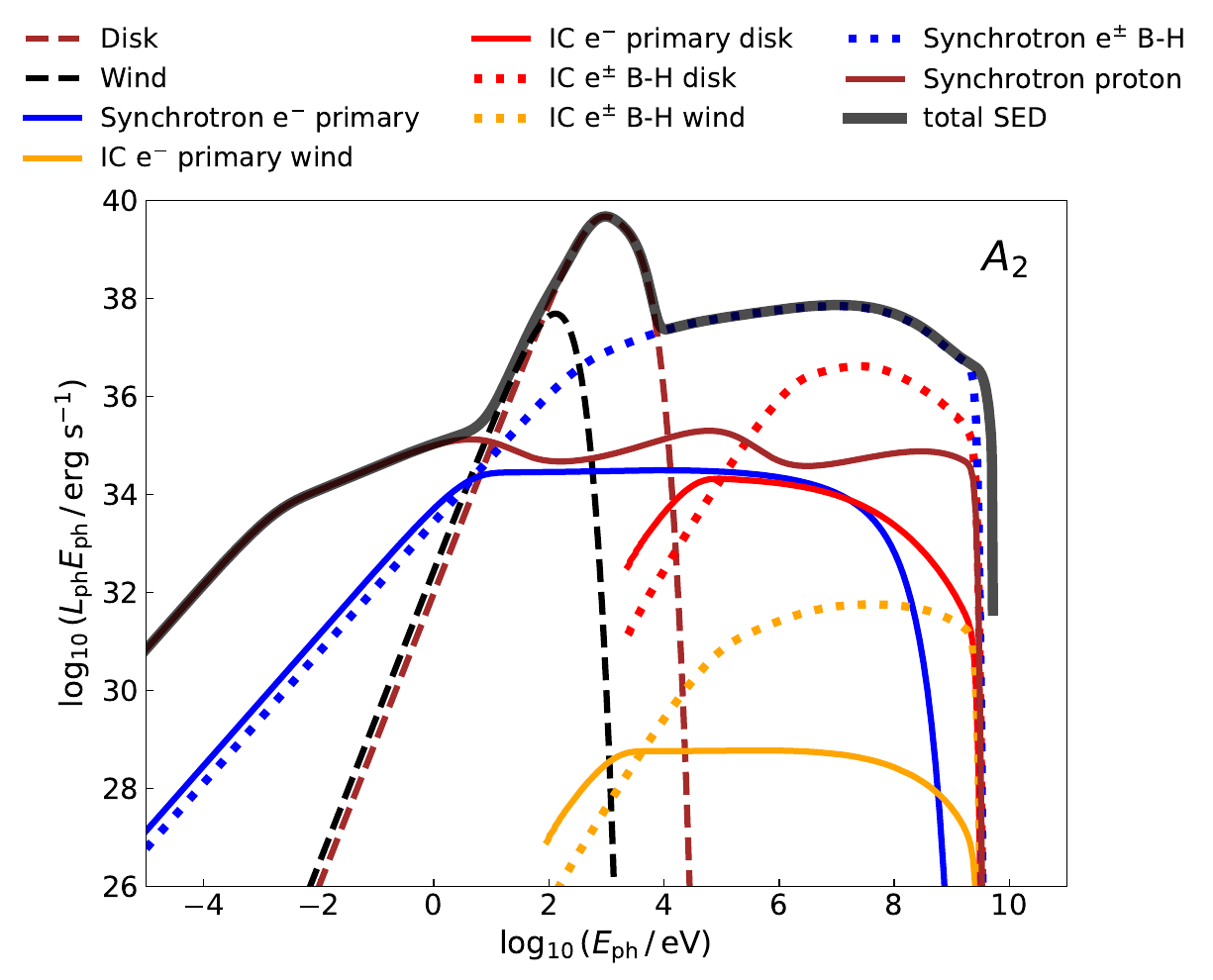}}
       \subfloat{
      \includegraphics[scale=0.4]{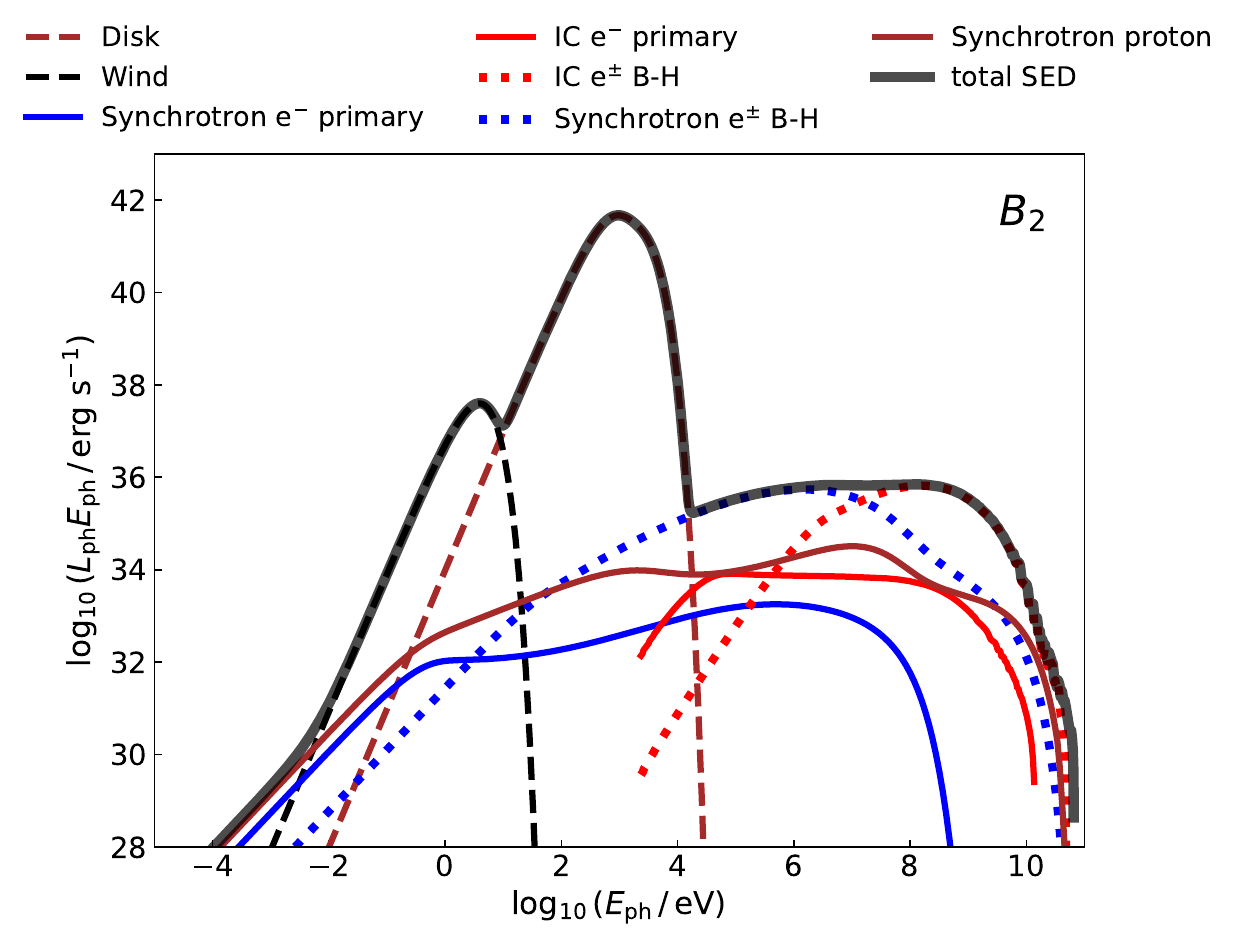}}
\caption{\small Similar to Fig. \ref{fig:seds} but for the scenario of particle acceleration by magnetic reconnection.}
\label{fig:seds_2}
\end{figure*}

Figure \ref{fig:seds_2} shows the SEDs of scenarios $A_2$ (left) and $B_2$ (right). In both cases, synchrotron and IC radiation from Bethe-Heitler secondary pairs are the dominant nonthermal processes. In scenario $A_2$, synchrotron radiation from secondary pairs dominates the high-energy spectrum, with a maximum luminosity of approximately $10^{38}\,{\rm erg\,s^{-1}}$, and a cut-off around $5\,{\rm GeV}$ due to $\gamma\gamma$ annihilation. In scenario $B_2$, the maximum luminosities of synchrotron and IC radiation from secondary pairs is $\sim 5 \times 10^{35}\,{\rm erg\,s^{-1}}$ around MeV and GeV. This emission surpasses that of the primary electrons across most of the energy range.

The absorption is calculated as in \cite{Cerutti-Dubus-CygX3_2011}. Absorption of nonthermal radiation produced by photon-photon annihilation is only relevant for energies higher than $\sim 10\,{\rm GeV}$ (scenarios $A_2$ and $B_2$) and $100\,{\rm GeV}$ (scenarios $A_1$ and $B_1$). We consider the thermal radiation fields from the disk and the photosphere of the wind as targets of the nonthermal photons. We find that emission produced by the photomeson mechanism is completely suppressed.

We also show the thermal SEDs of the disk-driven wind and the disk (in the observer's frame). As expected, the radiation of the disk dominates the ULX emission, with its maximum in the X-ray band, reaching apparent luminosities of $\sim 10^{40}\,{\rm erg\,s^{-1}}$ ($A$) and $\sim 10^{42}\,{\rm erg\,s^{-1}}$ ($B$). 
\begin{comment}
  In addition, we plot the sensitivity curves of different instruments for a distance of $\sim50\,$kpc, corresponding to the Magellanic Cloud: \textit{XMM-Newton} ($10^5$~s), \textit{Chandra} ($10^5$~s), \textit{NuSTAR} ($10^6$~s), \textit{Fermi}-LAT (10 years) where the integration time is given in brackets. In the case of \textit{e-ASTROGAM}, the sensitivity shown is for a $3\sigma$ detection after one year of effective exposition. %  
  (\citealt{2018JHEAp..19....1D}, \citealt{2022JCAP...08..013L}). 
\end{comment}

%\subsection{Application to NGC 4190 ULX-1}

We now apply our model to the source NGC 4190 ULX 1. This source has been classified as a hard ultraluminous due to the spectral hardness, which is consistent with a low inclination angle $i\approx0$, indicating a face-on orientation \citep{2021MNRAS.504..974G,Abaroaetal2023, Combi_etal2024}.

Following \cite{Abaroaetal2023}, we assume that the BH has a mass of $10M_{\odot}$ and accretes matter from the donor star at a rate of $\dot{M}=10\dot{M}_{\rm Edd}$. The fundamental disk parameters are the same as those listed in Table \ref{table: 1}.

\begin{figure}[h!]
    \centering   \includegraphics[width=0.9\linewidth]{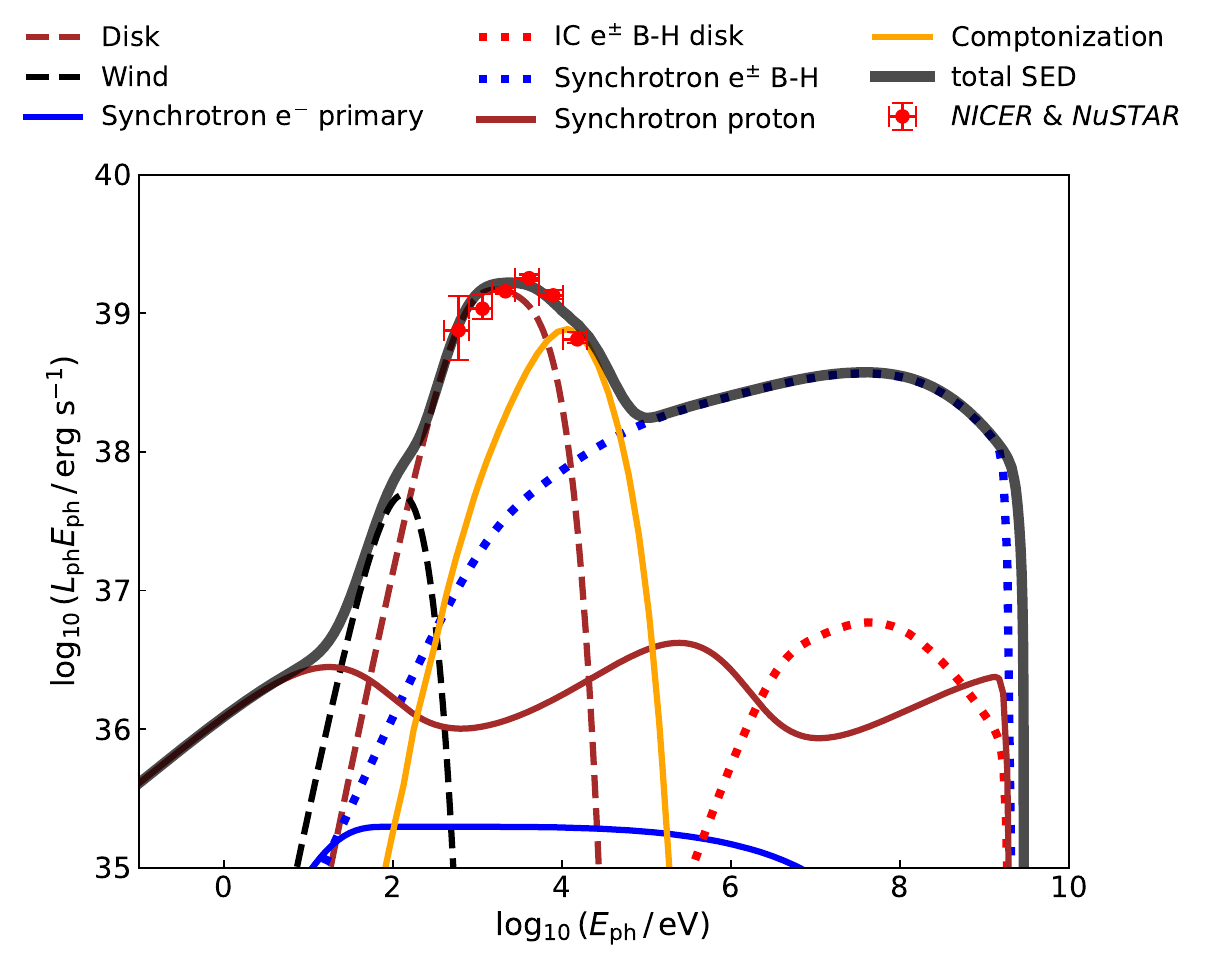}
    \caption{SED of the source NGC 4190 ULX-1, shown on a logarithmic scale. We show the thermal emission from the wind and innermost region of the accretion disk that escapes through the funnel, as well as non-thermal radiation produced by relativistic particles accelerated via magnetic reconnection. We add the inverse Compton SED from primary electrons accelerated far from the BH, following \cite{Combi_etal2024}. The red data points correspond to observations from \textit{NICER} and \textit{NuSTAR}. The solid black line is the total emission.} 
    \label{fig:SED_DATA}
\end{figure}

Figure \ref{fig:SED_DATA} shows observational data from \textit{NuSTAR} and \textit{NICER} (from \citealt{Combi_etal2024}) and the SEDs. Our model predicts the expected flux in the MeV–GeV range resulting from the interaction of secondary pairs with the magnetic field. In this energy range, the luminosity is $\sim 2\times  10^{38}\,{\rm erg\,s^{-1}}$. We remark the change in the slope at $\sim 100\,$ keV due to the pair contribution.This feature can be used to test the presence of relativistic protons in the source through future observations by \textit{COSI} satellite \citep{cosi2023}. %The main differences between our results and those from \cite{Combi_etal2024} are related to two parameters: the magnetization of the disk (a low-magnetized disk in our model, a non-magnetized disk in theirs) and the hadron-to-lepton ratio for the power of relativistic particles (a hadronic-dominated scenario in our model, a leptonic-dominated scenario in theirs). In the case of thermal emission, the introduction of magnetization in the disk slightly reduces the emission at hard X-rays. Regarding nonthermal emission, our model includes the study of the magnetic reconnection near the BH under a hadronic-dominated scenario, because we are mainly interested in the contribution of secondary pairs to the emission.}

\section{Gamma-ray absorption in the wind} \label{sec: opacity}

When the inclination angle of the system exceeds the semi-opening angle of the funnel ($i > \vartheta$), the radiation produced within the funnel is absorbed by the surrounding opaque wind. The gamma rays interact with the wind through Bethe-Heitler pair production and photomeson production: $\gamma + N \rightarrow N + e^- + e^+$ and $\gamma + N \rightarrow N + \pi^i$, where $N$ represents a nucleon.

We assume that the wind is smooth and that it expands isotropically at a constant velocity, its density can be expressed as \citep{Abaroa&Romero2024smbh}:
\begin{equation}
    \rho_{\rm w} = \frac{\dot{M}_{\rm w}}{4\pi r^2 v_{\rm w}},
\end{equation}
where $\dot{M}_{\rm w} \approx \dot{M}_{\rm input}$ is the wind mass-loss rate, and $r$ is the radial distance from the BH. The optical depth due to $\gamma$-nucleon interactions in the wind is given by \citet{2008MNRAS.387.1745R} as:

\begin{equation}
\tau_{\gamma N}(\vec{z}_j) = \int_0^{\infty} \sigma_{\gamma N} \left( \frac{\rho_{\rm w}}{m_{\rm p}} \right) d\rho_{\gamma},
\end{equation}
where $\sigma_{\gamma N} = \sigma_{\rm p\gamma}^{\pi^i} + \sigma_{\rm p\gamma}^{e^\pm}$ is the total cross section for $\gamma$-ray interactions with matter. The cross section parametrizations follow \citet{Maximon1968} and \citet{kelner2008}.

\begin{figure*}[h!]
    \centering
        \includegraphics[scale=0.45]{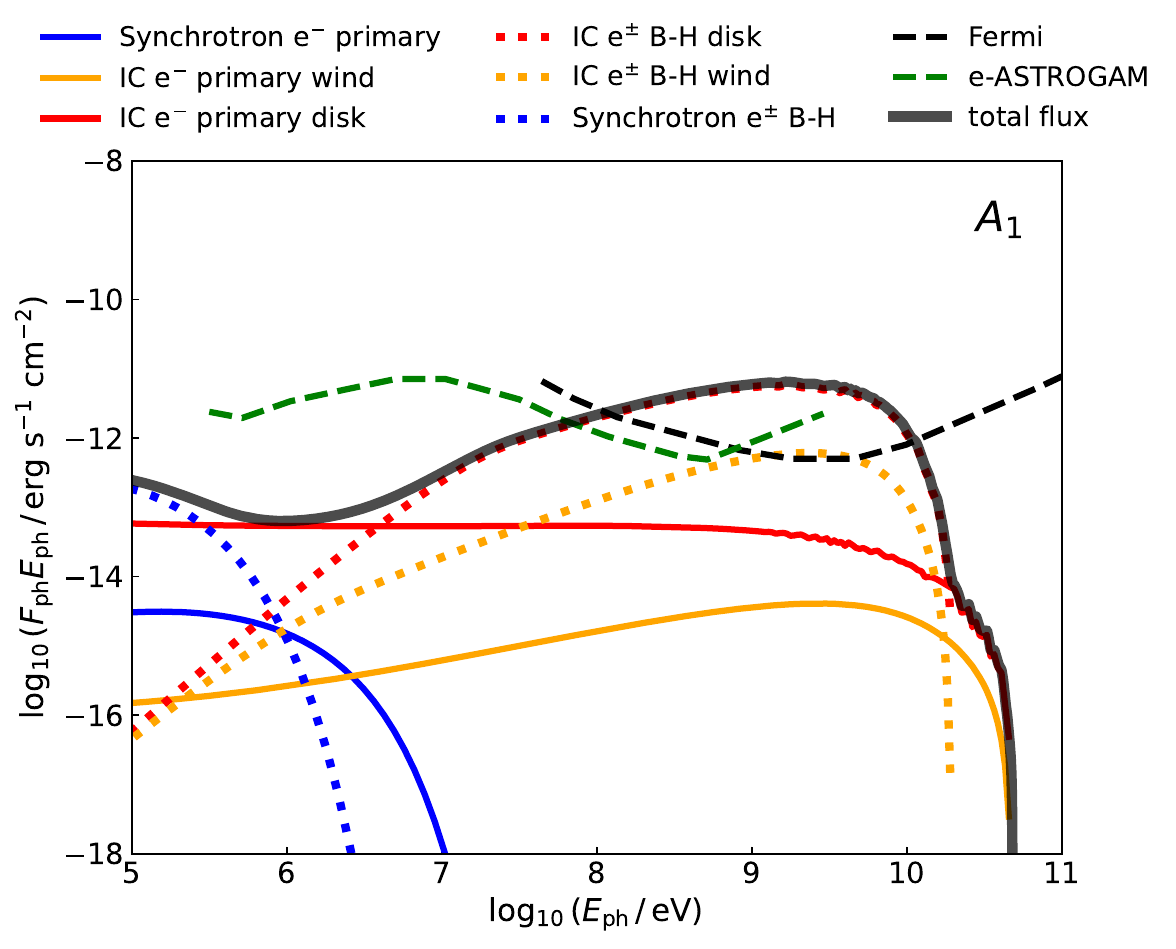}%
    \quad
{%
        \includegraphics[scale=0.45]{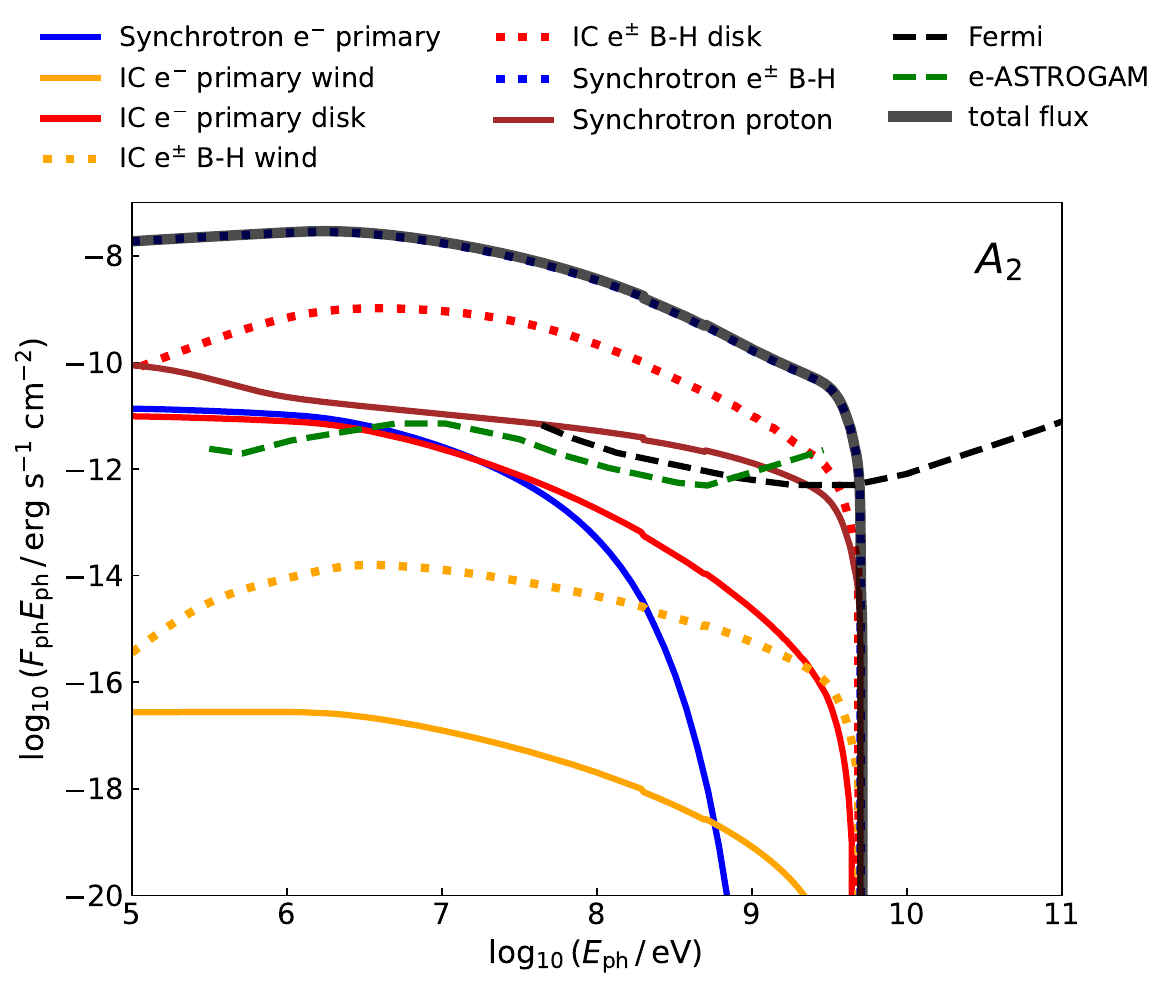}%
    }
    \caption{\small Predicted gamma-ray flux for a ULX with an inclination angle of $i = 45^\circ$ and a distance of $d = 15$\,kpc. Sensitivities of the space-based telescopes \textit{Fermi} and \textit{e-ASTROGAM} are shown with dashed lines. The solid black line is the overall emission.}
    \label{fig:flux}
\end{figure*}

Figure \ref{fig:flux} shows the gamma-ray flux for a galactic ULX located at 15 kpc and viewed at an inclination angle of $i = 45^\circ$, for scenarios $A_1$ and $A_2$, both corresponding to mass accretion rates of $\dot{m} = 10$. A fraction of the gamma radiation is attenuated by the wind, and above 10\,GeV, the absorption becomes total.

In scenarios with higher accretion rates, the wind becomes denser and can extend up to $\sim 2 \times 10^{12}\,$cm, leading to complete suppression of radiation above $\sim 1$\,MeV.

It is possible that some misaligned ULXs may exist within the Galaxy. These systems would remain undetected in X-rays due to wind obscuration, yet they could be observable in the MeV-GeV range if the accretion rates remain moderately super-Eddington.

\section{Discussion}
\label{sec:discussion}

The primary distinction between accretion scenarios $A$ ($\dot{m}_{\rm input} = 10$) and $B$ ($\dot{m}_{\rm input} = 10^3$) arise from properties that are highly sensitive to the accretion rate, such as the strength of the disk's magnetic field, the semi-opening angle of the funnel (which narrows with increasing accretion rate), and the geometric and physical characteristics of the wind.

Key parameters governing the disk's internal structure include the adiabatic index ($\gamma$), the advection parameter ($f$), the viscosity coefficient ($\alpha$), and the degree of magnetization of the plasma ($\beta$). In our modeling, we adopted intermediate values for these quantities to provide a representative overview of the physical behavior of such systems. Nevertheless, variations in these parameters would affect essential quantities—most notably, the toroidal component of the magnetic field ($B_\phi$)—and thereby influence the conditions in the particle acceleration region. The analytical framework presented here allows for straightforward recalculation under alternative assumptions, enabling researchers to explore a broader range of configurations. We consider the results reported in this work to be representative of the average behavior of these sources.

Another key factor is the location of the acceleration region for the nonthermal particle population. If this region lies farther from the BH, the disk radiation available for the Bethe-Heitler process diminishes significantly. As a result, the synchrotron emission from the resulting secondary pairs is also reduced. Moreover, the magnetic field typically decays with a toroidal profile beyond the Alfvén surface, which is generally located within $\sim$150 $r_{\rm g}$. As the field strength declines, both synchrotron losses and the particle acceleration rate decrease, leading to lower maximum energies for relativistic electrons and protons. Although the precise location of internal shock formation remains uncertain, it is plausible that shocks arise from local perturbations induced by the surrounding wind once the outflow becomes compressible.

In scenarios $A_1$ and $B_1$, we computed the SEDs assuming the maximum gas injection rate ($\dot{M}_{\rm gas}$) permitted within the funnel, thereby providing upper limits on the nonthermal radiation produced in these systems. However, under more realistic conditions, the actual number of particles entering the funnel is expected to be lower. Future multiwavelength observations—particularly in the X-ray and gamma-ray bands— could help to constrain the relevant parameter space by placing upper limits on the nonthermal luminosity, thereby providing indirect estimates of the power carried by relativistic particles, the funnel density, and the effective accretion rate. In addition, variability timescales and spectral slopes could offer clues about the location and efficiency of particle acceleration.

For scenarios $A_2$ and $B_2$, in which magnetic reconnection is the dominant acceleration mechanism, we adopted densities that yield an injected particle spectrum with index $p = 2$: specifically, $\rho_{A_2} = 0.1\rho_{\mathrm{max}}$ and $\rho_{B_2} = 0.001\rho_{\mathrm{max}}$. Deviations from these assumed densities would produce harder or softer particle spectra, significantly affecting the resulting emission. Thus, detailed spectral modeling and observational constraints will be essential for narrowing down the viable physical conditions in these sources.

This work does not explicitly model the mechanism by which matter is injected into the funnel. One plausible scenario involves the entrainment of hot plasma bubbles originating in the surrounding wind, which subsequently mix into the funnel region. Numerical simulations of supercritical accretion flows have shown that the accretion of complex, non-dipolar magnetic field configurations can lead to such outcomes by channeling gas into the polar regions of the BH \citep{McKinney2009,Romero2021AN}. These studies highlight the importance of magnetic field geometry in regulating mass loading and determining the internal structure of the funnel, which in turn influences the environment where particle acceleration occurs.

While such simulations provide valuable insights, they are limited by computational constraints and simplifying assumptions, particularly regarding radiative transfer and magnetic reconnection physics. Observational data, such as high-resolution imaging of jets and outflows or spectral signatures of entrained plasma, could help refine these models and provide empirical support for specific injection scenarios.

Our model predicts radiation emission in the largely unexplored energy range of 0.1-10 MeV from ULXs powered by BHs. In some extragalactic sources, such as NGC 4190 ULX-1, \textit{NuSTAR} has detected emission up to energies of $\sim$30 keV \citep{Combi_etal2024}, which has been interpreted as IC upscattering of disk photons. Our model offers an alternative explanation based on the presence of relativistic protons. These two interpretations can be observationally distinguished, as the Bethe-Heitler mechanism produces a population of secondary pairs whose emission should extend well into the MeV regime. Consequently, high-sensitivity observations in the MeV band—which have not been possible since the decommissioning of \textit{COMPTEL} on board the Compton Gamma Ray Observatory—are essential to probe the presence of relativistic protons in ULXs.

The proposed \textit{e-ASTROGAM} mission \citep{DEANGELIS20181}, currently under consideration by the European Space Agency (ESA), is designed to explore the gamma-ray sky in the 0.3 MeV to 3 GeV range with unprecedented sensitivity and polarization capability. Its payload will include a silicon tracker to detect Compton scattering and pair production events, along with a calorimeter to measure the energy of the secondary particles. \textit{e-ASTROGAM} has been proposed as the fifth medium-size (M5) mission in ESA’s Cosmic Vision program, with a potential launch date in 2029. Another highly relevant mission is \textit{COSI} (the Compton Spectrometer and Imager) \citep{2014cosp...40E3371T, cosi2023}, scheduled for launch in 2027. \textit{COSI} will operate in the 0.2-7 MeV band and is expected to provide critical observations to test scenarios involving hadronic components in ULXs.

Observations with next-generation MeV telescopes may also enable the detection of Galactic ULXs whose X-ray emission is heavily absorbed by the dense wind of the system when viewed at high inclination angles, while their softer gamma-ray emission—produced by secondary pairs—remains detectable. Since such misaligned systems are far more numerous than those viewed face-on, it is plausible that a significant population of obscured ULXs exists within our Galaxy. Their MeV radiation could serve as a key observational signature for their identification. Moreover, the model presented here could be tested in other super-Eddington accreting sources with jets, such as SS433 and Cyg X-3, where hadronic processes may also play a significant role in the high-energy emission.

\section{Conclusions}
\label{sec:conclusions}

We have investigated the consequences of relativistic proton injection into the funnels of ULXs powered by BHs. Our results show that protons can be accelerated to energies of $\sim$10-200 TeV and up to $\sim$5 PeV, depending on the accretion scenario. These protons interact with thermal X-band photons from the disk via the Bethe-Heitler mechanism, producing secondary electron-positron pairs. The resulting pairs radiate primarily through synchrotron and IC processes, generating nonthermal emission that extends from $\sim$1 MeV to $\sim$10 GeV, with luminosities in the range of $10^{33}$-$10^{38}$ erg s$^{-1}$.

We also find that photon-photon annihilation with the thermal radiation fields of the disk and wind does not significantly affect photons below $\sim$10 GeV. However, it effectively suppresses the gamma-ray emission produced by neutral pion decay from photomeson interactions. This reinforces the role of Bethe-Heitler pair production as the dominant channel for high-energy radiation in these systems.

An important implication of our results is the potential for detecting obscured ULXs within the Galaxy. While X-ray emission can be heavily absorbed by the dense, radiation-driven wind, particularly in systems viewed at high inclination angles, the softer gamma-ray radiation from secondary pairs may still escape and be detectable. This opens a promising avenue for identifying hidden populations of ULXs through their MeV emission. Although no current missions operate in this critical energy range, upcoming instruments such as \textit{e-ASTROGAM} and \textit{COSI} are well suited to search for this predicted component in nearby sources. Additionally, \textit{NuSTAR} observations in the tens-of-keV band could help identify candidate sources by revealing power-law spectra without an exponential cutoff, which may point to hadronic activity.

Finally, we emphasize that relativistic protons can produce high-energy signatures in BH binaries even in the absence of jets. The detection of such spectral features in the MeV band remains a major observational challenge, but it offers a compelling opportunity to probe the presence of hadronic processes and the physical conditions in supercritical accretion flows.

\section*{Acknowledgements}
This research was funded by PID2022-136828NB-C41/AEI/10.13039/501100011033/ and
through the “Unit of Excellence María de Maeztu 2020-2023” award to the Institute of Cosmos
Sciences (CEX2019-000918-M). LA thanks the Universidad Nacional de La Plata. Additional support came from PIP 0554 (CONICET).

%\appendix
%\section{Appendix title 1}

\bibliographystyle{elsarticle-harv} 
\bibliography{main}



%% For example:

%% \bibitem[Aladro et al.(2015)]{Aladro15} Aladro, R., Martín, S., Riquelme, D., et al. 2015, \aas, 579, A101

%%\end{thebibliography}
\end{document}